\documentclass[aps,reprint,superscriptaddress,nofootinbib,preprintnumbers,longbibliography]{revtex4-2}

\usepackage{url}
\usepackage{upgreek,amssymb,multirow}
\usepackage{graphicx,color}
\usepackage{mathtools}
\usepackage{float}
\usepackage{mciteplus}
\usepackage{blindtext}
 
\usepackage{amsmath}
\usepackage{extarrows}
\usepackage{braket}
\usepackage{tikz}
\usepackage{tikz-cd}

\usepackage{breakurl}
\usepackage[hyperindex,breaklinks]{hyperref}

\newcommand{\ie}{\begin{equation}\begin{aligned}}
\newcommand{\fe}{\end{aligned}\end{equation}}

\newcommand{\D}{\mathsf{D}}
\newcommand{\V}{\mathsf{V}}

\newcommand{\bZ}{{\mathbb{Z}}}

\newcommand{\ee}{\mathrm{e}}
\newcommand{\oo}{\mathrm{o}}

\begin{document}
  
\preprint{YITP-SB-2024-05}

\title{Cluster state  as a non-invertible symmetry protected topological phase}
 
\author{Sahand Seifnashri}

\affiliation{School of Natural Sciences, Institute for Advanced Study, Princeton, NJ} 

\author{Shu-Heng Shao}
\affiliation{C.\ N.\ Yang Institute for Theoretical Physics, Stony Brook University, Stony Brook, NY}

\begin{abstract}
We show that the standard 1+1d $\mathbb{Z}_2\times \mathbb{Z}_2$ cluster model has a non-invertible global symmetry, described by the fusion category Rep(D$_8$). 
Therefore, the cluster state is not only a $\mathbb{Z}_2\times \mathbb{Z}_2$ symmetry protected topological (SPT) phase, but also a non-invertible SPT phase.  
We further find two new commuting Pauli Hamiltonians for the other two Rep(D$_8$) SPT phases on a tensor product Hilbert space of qubits, matching the classification in field theory and mathematics.  
We identify the edge modes and the local projective algebras at the interfaces between these non-invertible SPT phases. 
Finally, we show that there does not exist a symmetric entangler that maps between these distinct SPT states.
\end{abstract}


\maketitle

\tableofcontents

\section{Introduction}

Symmetry protected topological (SPT) phases  \cite{Pollmann:2009mhk,Chen:2010gda,Chen:2010zpc,Fidkowski:2010jmn,Schuch:2011niz,Chen:2011pg} are some of the most fundamental quantum phases of matter.  
Without imposing any global symmetry, these phases are gapped with a unique, non-degenerate ground state, and are completely featureless. 
However, they become distinct topological states when we impose a  symmetry $G$, in the sense there cannot be a continuous $G$-symmetric deformation connecting these states without a phase transition.  
See \cite{Senthil:2014ooa,Chiu:2015mfr,Zeng:2015pxf} for  reviews.

The simplest example of an SPT phase is the 1+1d cluster Hamiltonian  \cite{son2012topological,2006RpMP...57..147N}:
\ie
H _\text{cluster}= -  \sum_{j=1}^{L} \, Z_{j-1}X_j Z_{j+1} \label{cluster.H} \,.
\fe
We assume the space is a closed periodic chain of $L$ qubits with even $L$. 
It has a  $\mathbb{Z}_2^\ee \times\mathbb{Z}_2^\oo$ symmetry generated by:
\ie\label{Z2Z2}
\eta^{\ee} = \prod_{j:\text{even}} X_j\,,~~~~
\eta^{\oo} = \prod_{j:\text{odd}} X_j\,.
\fe
The Hamiltonian is gapped with a unique ground state, known as the cluster state, satisfying $Z_{j-1} X_j Z_{j+1}\ket{\text{cluster}} = \ket{\text{cluster}}$.
Indeed, it is a commuting projector of Pauli operators with no relation, and therefore the ground state degeneracy is one. 

The cluster state is in a distinct $\bZ_2^\ee\times \bZ_2^\oo$ SPT phase compared to the product state $\ket{++\cdots +}$ (which is the ground state of the trivial Hamiltonian $H_\text{trivial} = -\sum_{j=1}^L X_j$).\footnote{Our convention is that  $X\ket{\pm}=\pm \ket{\pm}$, $Z \ket{0} = \ket{0}, Z \ket{1} =-\ket{1}$.}
Explicitly, it is given by \ie
\ket{\text{cluster}} = \V\ket{++\cdots +},~~
\text{where} ~\V = \prod_{j=1}^L \mathrm{CZ}_{j,j+1} \,. \label{cluster.state}
\fe
Here $\mathrm{CZ} _{j,j+1} = {1+Z_j +Z_{j+1}-Z_j Z_{j+1} \over2}$. 
Since $\V$, known as the cluster entangler, is a finite-depth circuit, the cluster state is in the same phase as the product state if we do not impose any global symmetry. 
However, if we impose the $\mathbb{Z}_2^\ee\times\mathbb{Z}_2^\oo$ symmetry, the cluster state is a distinct SPT state compared to the (trivial) product state \cite{Zeng:2015pxf}. 
Indeed, while $\V$ globally is  $\mathbb{Z}_2^\ee\times\mathbb{Z}_2^\oo$ symmetric, the individual gates $\mathrm{CZ}_{j,j+1}$ are not.

In recent years, the notion of global symmetry has been generalized in many different directions \cite{Gaiotto:2014kfa}, leading to many new topological phases. 
In particular, there has been a lot of progress on a novel kind of symmetry, known as non-invertible symmetries, in quantum field theory and condensed matter theory.
Non-invertible symmetries are implemented by conserved operators without inverses, and therefore are not described by group theory. 
Nonetheless, they lead to new selection rules, new constraints on the phase diagram, and more. 
See  \cite{McGreevy:2022oyu,Cordova:2022ruw,Brennan:2023mmt,Schafer-Nameki:2023jdn,Shao:2023gho} for recent reviews.

The critical Ising lattice model serves as the prototypical example of a gapless system with a non-invertible symmetry, associated with the Kramers-Wannier duality \cite{Grimm:1992ni,Oshikawa:1996dj,Aasen:2016dop,Aasen:2020jwb,Li:2023ani,Seiberg:2023cdc,Seiberg:2024gek}.  
However, this non-invertible symmetry has a Lieb-Schultz-Mattis-type constraint, implying that it is incompatible with a unique gapped ground state \cite{Seiberg:2024gek}.\footnote{For non-invertible symmetries in continuum field theory, the incompatibility  with a non-degenerate gapped phase is often defined as a generalized 't Hooft anomaly \cite{Chang:2018iay,Thorngren:2019iar,Zhang:2023wlu,Cordova:2023bja,Antinucci:2023ezl}. See  \cite{Komargodski:2020mxz,Choi:2023xjw,Choi:2023vgk} for the relation between this definition and the obstruction to gauging the non-invertible global symmetry. } 
It is then natural to ask what is the simplest non-degenerate gapped  phase protected by a non-invertible symmetry.

In this paper we show that the standard cluster Hamiltonian \eqref{cluster.H} has a non-invertible symmetry. Therefore,  the cluster state is not only a $\bZ_2^\ee \times\bZ_2^\oo$ SPT phase, but also a \textit{non-invertible SPT phase}, i.e., a non-degenerate gapped phase invariant under a non-invertible symmetry. 
We furthermore find two other distinct SPT states protected by the same non-invertible symmetry, matching the expectation from category theory \cite{tambara2000representations} and field theory \cite{Thorngren:2019iar}.

\section{Non-invertible symmetry of the cluster model}

The key observation is that the cluster Hamiltonian is invariant under the transformation\footnote{Throughout the paper, we use $\rightsquigarrow$ to denote a transformation implemented by a non-invertible operator, and $\mapsto$ for a conventional symmetry transformation implemented by a unitary operator.}
\ie
X_j \rightsquigarrow Z_{j-1} Z_{j+1}\,, \qquad Z_{j-1}Z_{j+1} \rightsquigarrow X_j  \,, \label{KW2}
\fe
which defines an automorphism of the algebra of $\bZ^\ee_2\times \bZ^\oo_2$ invariant operators. 
However, suppose this transformation were implemented by an invertible operator $U$ such that $UX_j U^{-1} = Z_{j-1}Z_{j+1}$, then $U\eta^\ee U^{-1} =\prod_{j:\text{even}}Z_{j-1}Z_{j+1} =1$, which is a contradiction. 
Instead, this transformation  is implemented by the following conserved operator
\ie
\D  =  T \D^\ee \D^\oo\,, \label{D.TDD}
\fe
where
\ie
~&\D^\ee = e^{2\pi i L\over16}
{1+\eta^{\ee}\over\sqrt{2}} {1-i X_L \over \sqrt{2} } \cdots {1- i Z_4 Z_2\over \sqrt{2} }
{1-i X_{2}\over\sqrt{2}} \,,\\
&\D^\oo=  e^{2\pi i L\over16}
{1+\eta^{\oo}\over\sqrt{2}} {1-i X_{L-1} \over \sqrt{2} }\cdots  {1- i Z_3Z_1\over \sqrt{2} }{1-i X_{1}\over\sqrt{2}} \,, \label{Do.De}
\fe
are the Kramers-Wannier operators on the even and odd sites in \cite{Seiberg:2023cdc,Seiberg:2024gek}.
Here $T$ is the lattice translation by one site which acts on local operators as $T X_j  T^{-1} = X_{j+1},\,TZ_j T^{-1} = Z_{j+1}$ and satisfies $T^{L}=1$. We have $T \D^\ee = \D^\oo T$ and $T \D  =\D T$.  
However, $\D$ is not a conventional symmetry operator because it has a kernel --- it annihilates every state that is not $\mathbb{Z}_2^\ee \times\mathbb{Z}_2^\oo$ symmetric. 
It implements \eqref{KW2} in the sense that
\ie\label{Daction}
	\D X_j = Z_{j-1}Z_{j+1}\D \,,~~~ \D Z_{j-1}Z_{j+1}  = X_j \D\,.
\fe

The operators $\eta^\ee$, $\eta^\oo$, and $\D$ commute with the Hamiltonian and satisfy the following algebra
\ie
&\D^2  = 1+\eta^{\ee} +\eta^{\oo}+\eta^{\ee}\eta^{\oo}\,,\\
&\eta^{\ee} \D = \D\eta^{\ee}=  \eta^{\oo} \D =\D\eta^{\oo} = \D\,. \label{ni.symm}
\fe
Note that even though the definition of $\mathsf{D}$ appears to involve a lattice translation $T$, the factors $\mathsf{D}^\ee,\mathsf{D}^\oo$ each involves a half translation in the opposite direction, and hence the algebra of $\mathsf{D}$ does not mix with lattice translations.
We conclude that the cluster state is a topological phase protected by a non-invertible symmetry.

In Appendix \ref{app:D8}, we show that this non-invertible symmetry is described by the fusion category Rep(D$_8$), whose fusion algebra is given by the tensor product of the irreducible representations of the group D$_8$.
(See \cite{etingof2016tensor} for a review of fusion category, which is built on  \cite{Moore:1988qv,Moore:1988ss}.)
We prove this by gauging the non-invertible symmetry and finding a dual D$_8$ symmetry.

Finally, we note that the cluster Hamiltonian is also invariant under another non-invertible symmetry $\tilde \D=\D^\ee \D^\oo$, which obeys the algebra $\tilde\D^2 = (1+\eta^\ee +\eta^\oo +\eta^\ee \eta^\oo)T^{-2}$. 
Similar to the Kramers-Wannier symmetry of the critical Ising model \cite{Seiberg:2023cdc}, $\tilde\D$  mixes with the lattice translation, therefore it does not form a fusion category on the lattice. In the continuum, it can flow to a Rep(H$_8$) fusion category symmetry. 
Note that breaking the translation symmetry $T$ in \eqref{cluster.H} also breaks $\tilde \D$, but preserves $\D$.

\section{Non-invertible SPT phases}

We have shown that the cluster state is a non-invertible SPT phase. Are there other  Rep(D$_8$) SPT phases?
As we discuss in Section \ref{sec:entangler}, there is no notion of a trivial SPT phase for non-invertible symmetries. Also note that $\D$ does not act on-site and the product state is not invariant under it.\footnote{Even though non-invertible symmetries do not act on-site, they can sometimes leave the product state invariant. For instance, by performing the similarity transformation $\V$ we find the non-invertible symmetry operator $\V\D \V$ that does leave the product state invariant. See also \cite{Fechisin:2023dkj}, where the authors considered SPT phases protected by  the non-invertible symmetry $G\times \text{Rep}(G)$. In their lattice model, the product state is invariant under their non-invertible symmetry.}
Mathematically,  a non-invertible SPT phase corresponds to  a   fiber functor of the fusion category. 
It is known \cite{tambara2000representations} that there are 3 distinct fiber functors for Rep(D$_8$), labeled by the 3 non-trivial elements $\eta^\ee ,\eta^\oo, \eta^\text{d}=\eta^\ee\eta^\oo$ of $\mathbb{Z}_2^\ee \times\mathbb{Z}_2^\oo$. See  \cite{Thorngren:2019iar,Inamura:2021wuo,Cordova:2023bja,Bhardwaj:2023idu,Bhardwaj:2024qrf} for a field theory exposition of these SPT phases.

What are the lattice models for the other two Rep(D$_8$) SPT phases? To find them, we partially gauge the non-invertible symmetry and look for different spontaneous symmetry breaking patterns in the gauged theory. We first review this method in the case of $\mathbb{Z}_2^\ee\times \mathbb{Z}_2^\oo$.

\subsection{Gaugings and the KT transformation}

For $\mathbb{Z}_2^\ee\times \mathbb{Z}_2^\oo$, there are only two distinct SPT phases: the product state $\ket{++ \dots +}$ and the cluster state. 
One way to distinguish these two phases is to gauge the $\bZ^\ee_2\times \bZ^\oo_2$ symmetry. Gauging the symmetry via minimal coupling amounts to doing Kramers-Wannier transformations on even and odd sites separately. See Appendix \ref{app:S} for a review.  Up to a lattice translation, this is the same transformation as in \eqref{KW2}. The Hamiltonians of these two phases after gauging are 
\ie\label{S}
		H'_\text{trivial}  &= -  \sum_{j=1}^L\, Z_{j-1}' Z_{j+1}' \,,\\
	H'_\text{cluster}  &= -  \sum_{j=1}^L \, Z_{j-1}' X_j'  Z_{j+1}' \,.
	\fe 
The two SPT phases are distinct in that they are mapped to a symmetry breaking phase and a symmetry preserving phase under gauging. In particular, the cluster Hamiltonian is invariant under gauging.\footnote{Following the half-gauging argument in \cite{Chang:2018iay,Choi:2021kmx,Choi:2022zal,Seiberg:2024gek}, the invariance under gauging a global symmetry implies the existence of the non-invertible global symmetry, which is precisely the operator  $\D$. This will be explored in \cite{Seifnashri:2025fgd}.}

For later convenience, we will choose to distinguish these two SPT phases by a twisted gauging, reviewed in Appendix \ref{app:TST}. In the condensed matter literature, it is  known as the Kennedy-Tasaki (KT) transformation \cite{kennedy1992hidden,PhysRevB.45.304} (see also \cite{Li:2023ani,ParayilMana:2024txy}):
\ie
	X_j \rightsquigarrow \hat X_j \,, \qquad Z_{j-1}Z_{j+1} \rightsquigarrow \hat Z_{j-1}\hat X_j \hat Z_{j+1}  \,. \label{KT}
\fe
(In the field theory context, the untwisted and twisted gaugings are respectively referred to as the $\mathbf{S}$ and $\mathbf{TST}$ gaugings, representing different elements of the modular group.)
If we identify the hatted operators with the unhatted ones, then the KT transformation is implemented by the non-invertible operator $\V \D \V$ \cite{ParayilMana:2024txy}:
\ie
&(\V\D\V) X_j = X_j( \V\D\V) \,,\\
&(\V\D\V) Z_{j-1} Z_{j+1}  = Z_{j-1} X_j Z_{j+1} (\V\D\V)\,.
\fe 
The two SPT phases after twisted gauging become
\ie\label{TST}
	\hat{H}_\text{trivial}  &= -  \sum_{j}\, \hat X_j \,,\\
		\hat{H}_\text{cluster}  &= - \sum_{j}\, \hat Z_{j-1} \hat Z_{j+1} \,.
\fe
The trivial phase is invariant under twisted gauging and the cluster state goes to the symmetry breaking phase of the dual $\hat\bZ_2^\ee\times \hat \bZ_2^\oo$ global symmetry,  generated by $\hat\eta^{\ee}=  \prod_{j:\text{even}}\hat X_j, ~\hat \eta^{\oo} = \prod_{j:\text{odd}}\hat X_j$. 
See Table \ref{tab:Z2Z2} for a summary and  Appendix \ref{app:cont} for the (twisted) gauging the two SPT phases in  continuum field theory.

\begin{table}
\begin{tabular}{ ccc } 
\multirow{2}{10em}{$\bZ_2^\ee \times \bZ_2^\oo$ SPT states} &  & Symmetry breaking\\
&& pattern of  $\hat\bZ_2^\ee \times \hat\bZ_2^\oo$\\
\hline
$\ket{++ \cdots +}$ & \multirow{2}{4em}{ $\xrightarrow[~~\text{KT}~~]{\mathbf{TST}} $}  & unbroken \\ 
 $\ket{\text{cluster}}$ &  & completely broken 
\end{tabular}
 \caption{The KT transformation, corresponding to a twisted gauging labeled by $\mathbf{TST}$, maps the product state and the cluster state to a symmetry  preserving phase and a  symmetry breaking phase, respectively.}\label{tab:Z2Z2}
\end{table}

\subsection{Three Rep(D$_8$) SPT phases \label{sec:3.SPTs}}

Now we are ready to study the SPT phases of the non-invertible symmetry. As we show in Appendix \ref{app:t.gauging}, after the twisted gauging of $\bZ_2^\ee \times \bZ_2^\oo$, the non-invertible symmetry \eqref{ni.symm} becomes an anomalous $\hat{\bZ}_2^\ee \times \hat{\bZ}_2^\oo \times \hat{\bZ}_2^\V$ symmetry generated by
\ie
	\hat{\eta}^\ee = \prod_{j:\text{even}} \hat{X}_j \,, ~~ \hat{\eta}^\oo = \prod_{j:\text{odd}} \hat{X}_j\,, ~~ \hat{\V} = \prod_j \hat{\mathrm{CZ}}_{j,j+1}\,. 
\fe
In other words, applying the transformation \eqref{KT} to any Hamiltonian invariant under \eqref{ni.symm} results in a Hamiltonian with the $\hat{\bZ}_2^\ee \times \hat{\bZ}_2^\oo \times \hat{\bZ}_2^\V$ symmetry. The 't Hooft anomaly of this symmetry is described by the 2+1d invertible field theory 
$(-1)^{\int \hat{A}^\ee \cup \hat{A}^\oo \cup \hat{A}^\V} $, known as the type III anomaly \cite{deWildPropitius:1995cf,Wang:2014tia}.  
(See Appendix \ref{app:typeIII}. Here $\hat{A}^\ee$, $\hat{A}^\oo$, and $\hat{A}^\V$ are discrete background gauge fields.)
We summarize this relation as
\ie\label{D8typeIII}
\text{Rep(D$_8$)} \xlongrightarrow[\text{KT}]{\mathbf{TST}}
\underbrace{\hat \bZ_2^\ee \times \hat\bZ_2^\oo \times\hat \bZ_2^\V}_\text{type III anomaly}
\fe
The three non-invertible SPT phases can be distinguished by the symmetry breaking patterns of the this dual symmetry after the KT transformation. Equivalently, these  phases are distinguished by certain non-local, string order parameters (see, e.g., \cite{Bhardwaj:2023idu,Bhardwaj:2023fca}).

Given any of the three Rep(D$_8$) SPT phases, the fact that it has the non-invertible symmetry $\D$ implies the invariance under the untwisted gauging of $\bZ_2^\ee \times\bZ_2^\oo$  \cite{Seiberg:2024gek,Seifnashri:2025fgd}. As discussed above,  the cluster state is invariant under the untwisted gauging but the product state is not. Therefore, any  Rep(D$_8$) SPT  must be in the same phase as the cluster state as far as the $\bZ_2^\ee \times \bZ_2^\oo$ symmetry is concerned.

This implies that after the twisted gauging, the dual $\hat\bZ_2^\ee \times \hat\bZ_2^\oo$ symmetry must be completely broken, with a single $\bZ_2$ subgroup of $\hat\bZ_2^\ee \times \hat\bZ_2^\oo\times\hat\bZ_2^\V$  preserved.\footnote{The entire $\hat\bZ_2^\ee \times \hat\bZ_2^\oo\times\hat\bZ_2^\V$ cannot be completely broken. This would have led to 8-fold degenerate ground states which cannot arise from gauging $\bZ_2 \times \bZ_2$ of a model with a single gapped ground state.}
A priori, there are four options for this unbroken $\bZ_2$ subgroup,   generated by $\hat \V$, $\hat{\V}  \hat\eta^\ee$, $\hat{\V} \hat\eta^\oo$, or $\hat{\V} \hat\eta^\ee \hat\eta^\oo$. However, the diagonal subgroup generated by $\hat \V \hat\eta^\ee \hat\eta^\oo$ cannot be preserved since it is anomalous, which is obtained from $(-1)^{\int \hat{A}^\ee \cup \hat{A}^\oo \cup \hat{A}^\V} $ by setting $\hat A^\ee =\hat A^\oo =\hat A^\V$.
 This leaves us with three possible symmetry breaking patterns in Table \ref{tab:RepD8}.

\paragraph{$\hat \bZ_2^\V$ preserving phase:} The original cluster Hamiltonian after gauging is given by \eqref{TST}. The order parameters of the gauged theory are given by $\hat Z_0$ and $\hat Z_1$ which are both invariant under $\hat \V$. Thus the cluster state preserves the $\hat\bZ_2^\V$ subgroup.

\paragraph{$\text{diag}(\hat\bZ_2^\V \times \hat\bZ_2^\ee)$ preserving phase:} For the second option, we propose the following Hamiltonian for the gauged system\footnote{In our conventions for indices $j=1,\cdots, L$ and $n=1,\cdots, L/2$.}
\ie
 \hat{H}_\text{odd}& = \sum_{n=1}^{L/2} \hat  Z_{2n-1} \hat Z_{2n+1} \\
	&
-\sum_{n=1}^{L/2}  \hat Y_{2n} \hat Y_{2n+2}(1 +\hat Z_{2n-1} \hat Z_{2n+3}) \,,
\fe
and take the number of sites $L$ to be a multiple of 4. (The subscript ``odd" will become clear later.) This is a commuting  Pauli Hamiltonian  and thus is exactly solvable.\footnote{Crucially, the Hamiltonian is frustration-free in the sense that even though different terms in the Hamiltonian are not linearly independent,  it is consistent to minimize all of them simultaneously.} Note that the projection factor in the second term does not affect the ground states. This is because for the ground states $\hat Z_{2n-1}\hat  Z_{2n+1} = -1$ and thus $\hat Z_{2n-1}\hat Z_{2n+3} = 1$. The ground space is determined by $L-2$ independent constraints $\hat Z_{2n-1}\hat  Z_{2n+1}  = -1$ and $\hat Y_{2n} \hat Y_{2n+2}=1$. This leads to a four-fold degeneracy with ground states given by $\hat Z_1 = -\hat Z_3 = \cdots = - \hat Z_{L-1} = \pm 1$ and $\hat Y_0 =\hat  Y_2 = \cdots = \hat Y_{L-2} = \pm 1$.\footnote{We have used the fact that   $L$ is a multiple of 4.} The two order parameters can be taken to be $\hat Z_1$ and $\hat Y_2(1 -\hat  Z_1\hat Z_3)$, which are both invariant under $\hat \V \hat \eta^\ee$. 

Next, we undo the twisted gauging of $\bZ_2^\ee \times \bZ_2^\oo$ to find the Hamiltonian for this Rep(D$_8$) SPT phase:
\ie
 H_\text{odd} =&\sum_{n=1}^{L/2}   Z_{2n-1} X_{2n} Z_{2n+1} 
-\sum_{n=1}^{L/2}  Y_{2n} X_{2n+1} Y_{2n+2}  \\
&+\sum_{n=1}^{L/2}  Z_{2n-1}Z_{2n}  X_{2n+1} Z_{2n+2}Z_{2n+3} \,. \label{o.SPT}
\fe
This Hamiltonian has a unique ground state, denoted by $\ket{\text{odd}}$, stabilized by the following $L$ generators:
\ie\label{oddstabilizer}
	-Z_{2n-1} X_{2n} Z_{2n+1} = 1 \,, \quad  Y_{2n} X_{2n+1} Y_{2n+2}  = 1 \,.
\fe
Note that while  $-Z_{2n-1} X_{2n} Z_{2n+1} $ is invariant under $\D$,   $ Y_{2n} X_{2n+1} Y_{2n+2} $ is mapped  to a product of terms in \eqref{oddstabilizer}.  
Similarly, the individual terms  in the first sum of $H_\text{odd}$ are invariant under $\D$, while the terms in the second and the third sums are mapped to each other. 

The $\ket{\text{odd}}$ state is related to the product state as
\ie
	\ket{\text{odd}}=\prod_{n=1}^{L/2} \mathrm{CZ}_{2n-1,2n+1} \prod_{j=1}^L \mathrm{CZ}_{j,j+1} \ket{-- \cdots -}\,.
\fe
In Appendix \ref{app:D.action}, we will show $\D \ket{\text{odd}} = 2(-1)^{L/4} \ket{\text{odd}}$, while $\D \ket{\text{cluster}} = 2\ket{\text{cluster}}$.

\paragraph{$\text{diag}(\hat \bZ_2^\V \times \hat\bZ_2^\oo)$ preserving phase:} The third option is related to the previous one by exchanging $\ee \leftrightarrow \oo$. More precisely, the Hamiltonian for this state, denoted as $\ket{\text{even}}$, is obtained by conjugating $H_\text{odd}$ with the lattice translation, i.e., $H_\text{even} = T H_\text{odd} T^{-1}$.

\begin{table}
\begin{tabular}{ ccc } 
\multirow{2}{10em}{Rep(D$_8$) SPT states}  &  & Symmetry breaking\\
&&  pattern of  $\hat\bZ_2^\ee \times \hat\bZ_2^\oo\times \hat \bZ_2^\V$\\
\hline
$\ket{\text{cluster}}$ & \multirow{3}{4em}{ $\xrightarrow[~~~\text{KT}~~~]{\mathbf{TST}} $}  &   $\hat\bZ_2^\V$ unbroken\\ 
 $\ket{\text{odd}}$ &  & $\text{diag}(\hat\bZ_2^\V \times\hat\bZ_2^\ee)$ unbroken \\
$ \ket{\text{even}}$ && $\text{diag}(\hat\bZ_2^\V \times\hat\bZ_2^\oo)$ unbroken  
\end{tabular}
  \caption{The three Rep(D$_8$) SPT phases are distinguished by the symmetry breaking patterns of the dual $\hat\bZ_2^\ee \times \hat\bZ_2^\oo\times \hat \bZ_2^\V$  symmetry after the KT transformation (i.e., a twisted gauging labeled by $\mathbf{TST}$). For all three phases, $\hat\bZ_2^\ee\times\hat \bZ_2^\oo$ is spontaneously broken, but the unbroken subgroup is different.}\label{tab:RepD8}
\end{table}

~\\

We have thus identified three Rep(D$_8$) SPT phases, $\ket{\text{cluster}}$, $\ket{\text{odd}}$, and $\ket{\text{even}}$.  
The SPT states $\ket{\text{odd}}$ and $\ket{\text{even}}$ are related to the cluster state by the   finite-depth circuits
\ie
	\prod_{j=1}^L Z_j \prod_{n=1}^{L/2} \mathrm{CZ}_{2n-1,2n+1} \quad \text{and} \quad \prod_{j=1}^L Z_j \prod_{n=1}^{L/2} \mathrm{CZ}_{2n,2n+2}\,, \label{circuits}
\fe
respectively. The gates of the circuits can be made $\bZ^\ee_2\times \bZ^\oo_2$-symmetric:
\ie
	\ket{\text{odd}} = \prod_{k=1}^{L/4} Z_{4k-2}Z_{4k} \prod_{n=1}^{L/2} e^{\frac{\pi i}{4} (Z_{2n-1}Z_{2n+1}-1) } \ket{\text{cluster}}\,. \label{circuit.o.d}
\fe
This shows that the three SPT states  are in the same $\bZ^\ee_2\times \bZ^\oo_2$ SPT phase. However, they are different SPT phases for Rep(D$_8$). Indeed the circuits in \eqref{circuits} do not commute with the non-invertible symmetry operator $\D$ (see Section \ref{sec:entangler} for more discussions).

\section{Edge modes}

In the previous section, we distinguished the three SPT phases by their different symmetry breaking patterns after gauging a part of the non-invertible symmetry. Here, we discuss the edge modes on the boundary of these phases.

For ordinary on-site symmetries, a (non-trivial) SPT phase is characterized by the edge modes or the projective algebra of the symmetry operator on an open chain with two boundaries. 
However, our non-invertible symmetry is not a product of local unitary operators, and it is unclear how to ``cut it open." 
We therefore need a more universal diagnostic for SPT phases, one that is equivalent to the standard one for on-site symmetries, but also applicable to symmetries that are not necessarily on-site. 
To this end, we reinterpret an open chain system as an \textit{interface} between the (non-trivial) SPT phase and the product state on a closed chain.  
The relative difference between this SPT phase and the product state is then diagnosed by the edge modes at the interfaces. 
The advantage of this simple reinterpretation is that we never have to cut open the symmetry operator, and this criterion can be readily generalized to non-invertible symmetries. 
See Appendix \ref{app:clusterprod} for a detailed discussion of this reinterpretation of the edge modes in the cluster model.

Here we focus on  the interface between the SPT states $\ket{\text{cluster}}$ and $\ket{\text{odd}}$. Consider a closed chain of $L$ sites, where the cluster state is on half of the chain between sites $1$ and $\ell$ with $1 < \ell < L$, and the $\ket{\text{odd}}$ state is in the region between sites $\ell$ and  $L$. We assume $L-\ell$ to be  a multiple of 4 and $L$ to be even.
Consider the Rep(D$_8$)-symmetric interface Hamiltonian:
\ie
	&H_{\text{cluster}|\text{odd}} = - \sum_{j=1}^\ell Z_{j-1}X_jZ_{j+1}+ \sum_{n={\ell\over2}+1}^{L \over 2} Z_{2n-1}X_{2n}Z_{2n+1} \\
	&- \sum_{n={\ell \over 2}+1}^{\frac L2-2} Y_{2n}X_{2n+1}Y_{2n+2}\left(1 + Z_{2n-1}X_{2n}X_{2n+2}Z_{2n+3} \right) \,. \label{interface}
\fe
The ground space ${\cal H}_{\text{cluster}|\text{odd}}$ is stabilized by the following $2L-2$ generators:
\ie
	Z_{j-1} X_j Z_{j+1}=1 \quad &\text{ for } j = 1, \cdots, \ell \,, \\
	-Z_{2n-1}X_{2n}Z_{2n+1} = 1 \quad &\text{ for } n = \ell/2 +1, \cdots ,L/2 \,, \\
	Y_{2n}X_{2n+1}Y_{2n+2} = 1 \quad &\text{ for } n = \ell/2 + 1, \cdots ,L/2 - 2 \,. \label{stabilizers}
\fe
Hence the ground space  is four-fold degenerate,  signaling edge modes at the interfaces.

\begin{figure}
\begin{tikzpicture}
\draw(0,0) ellipse (1 and 0.5);
\filldraw [black] (-.4,-0.46) circle (1pt)node [below] {\tiny $\ket{\text{odd}}$} ;
\filldraw [black] (0, -0.5) circle (1pt)   ;
\filldraw [black] (.4,-0.46) circle (1pt) ;
\filldraw [black] (.75,-0.33) circle (1pt) ;
\filldraw [black] (-.75,-0.33) circle (1pt) ;
\filldraw [black] (-1,0) circle (1pt)node[left]{\tiny $L$-$1$};
\filldraw [black] (-.75,0.33) circle (1pt)node[left]{\tiny $L$};
\filldraw [black] (-.4,0.46) circle (1pt)node  [above]{\tiny $1$} ;
\filldraw [black] (0, 0.5) circle (1pt) ;
\filldraw [black] (.4,0.46) circle (1pt) ;
\filldraw [black] (.75,0.33) circle (1pt) node  [right]{\tiny $\ell$};
\filldraw [black] (1,0) circle (1pt)node[right]{\tiny $\ell$+$1$};
\draw [black] (0,.65)  node [right] {\tiny $\ket{\text{cluster}}$};
\draw[red,thick ](0,1.3) ellipse (1 and 0.5);
\draw [red] (-1.4,1.3)  node[right]{ \small $\D$};
\end{tikzpicture}
\begin{tikzpicture}
\draw [black] (-3,0)  node[right]{  $=\sum\limits_{I=1,2}$};
\draw(0,0) ellipse (1 and 0.5);
\filldraw [black] (-.4,-0.46) circle (1pt)node [below] {\tiny $\ket{\text{odd}}$} ;
\filldraw [black] (0, -0.5) circle (1pt)   ;
\filldraw [black] (.4,-0.46) circle (1pt) ;
\filldraw [black] (.75,-0.33) circle (1pt) ;
\filldraw [black] (-.75,-0.33) circle (1pt) ;
\filldraw [red] (-1,0) circle (1pt)node[left]{\tiny $\D_\text{L}^{(I)}$};
\filldraw [black] (-.75,0.33) circle (1pt) ;
\filldraw [black] (-.4,0.46) circle (1pt) ;
\filldraw [black] (0, 0.5) circle (1pt) ;
\filldraw [black] (.4,0.46) circle (1pt) ;
\filldraw [black] (.75,0.33) circle (1pt) ;
\filldraw [red] (1,0) circle (1pt)node[right]{\tiny $\D_\text{R}^{(I)}$};
\draw [black] (0,.65)  node [right] {\tiny $\ket{\text{cluster}}$};
\end{tikzpicture}
\caption{Localization of the non-invertible operator  $\D$ at the interfaces between SPT phases.  The interface system \eqref{interface} is locally in the cluster state in the region $j=1,2,\cdots, \ell$, and locally in the $\ket{\text{odd}}$  state in the complement region. The operator $\D$ factorizes into local factors $\D_\text{L}^{(1)}\D_\text{R}^{(1)}+\D_\text{L}^{(2)}\D_\text{R}^{(2)}$ (multiplied by $(-1)^{(L-\ell)/ 4}$) on the ground space . }\label{fig:factorization}
\end{figure}
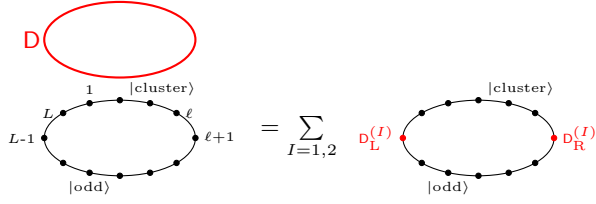

Before identifying the edge modes, we first discuss the projective algebra of the non-invertible symmetry at each of the interfaces, which protects the edge modes.
As shown in Appendix \ref{app:noninv.interface}, the action of  symmetry operators on any state $\ket{\psi} \in {\cal H}_{\text{cluster}|\text{odd}}$ factorizes into local factors at the interfaces:
\ie
	\D \ket{\psi} &= (-1)^{\frac{L-\ell}{4}} \left(\D_\mathrm{L}^{(1)}\D_\mathrm{R}^{(1)} + \D_\mathrm{L}^{(2)}\D_\mathrm{R}^{(2)}\right) \ket{\psi} \,, \\
	\eta^\oo \ket{\psi} &= \eta^\oo_\mathrm{L} \eta^\oo_\mathrm{R} \ket{\psi} \,, ~~~	\\
	\eta^\ee \ket{\psi}& =  \ket{\psi} \,,
\fe
where
\ie
	\D_\mathrm{L}^{(1)} &= Y_{L-2} Y_{L-1} Z_L \,,  ~&  \D_\mathrm{R}^{(1)} &= Z_{\ell + 1} \,, \\
	\D_\mathrm{L}^{(2)} &= Z_{L-1} \,,  &  \D_\mathrm{R}^{(2)} &= Z_\ell Y_{\ell+1} Y_{\ell +2} \,, \\
	\eta^\oo_\mathrm{L} &= Y_{L-2} X_{L-1} Z_L\,,  & \eta^\oo_\mathrm{R} &= Z_\ell X_{\ell+1} Y_{\ell +2} \,.\label{edge.modes}
\fe
Here L and R stand for the left and right interfaces around sites $j=L$ and $j=\ell$, respectively; see Figure \ref{fig:factorization}. 
We find that the \textit{local} factors of $\D$ are charged under $\eta^\oo$ (but not  under $\eta^\ee$):
\ie
	\eta^\oo \, \D_\mathrm{L}^{(I)} &= - \D_\mathrm{L}^{(I)} \, \eta^\oo \,,  ~& \eta^\oo \, \D_\mathrm{R}^{(I)} &= - \D_\mathrm{R}^{(I)} \,\eta^\oo \,,~~ I=1,2\,. \label{proj.alg}
\fe
Therefore, there is a projective algebra at each interface between the local factors:
\ie
\eta^\oo_\text{L} \D^{(I)}_\text{L} &= - \D^{(I)}_\text{L} \eta^\oo_\text{L} \,, ~&\eta^\oo_\text{R} \D^{(I)}_\text{R} &= -\D^{(I)}_\text{R} \eta^\oo_\text{R}  \,,~~I=1,2\,, \label{loc.alg}
\fe
whereas $\eta^\ee_\text{L} \D^{(I)}_\text{L} = \D^{(I)}_\text{L} \eta^\ee_\text{L}$ and $\eta^\ee_\text{R} \D^{(I)}_\text{R} = \D^{(I)}_\text{R} \eta^\ee_\text{R}$.
This projective algebra matches with the continuum discussion in \cite[(3.62)]{Thorngren:2019iar}.
Importantly,  the \textit{global} $\D, \eta^\ee,\eta^\oo$ operators realize the algebra \eqref{ni.symm} linearly.

On the left interface, the operators $\eta^\oo_\mathrm{L}$ and $\D_\mathrm{L}^{(2)}$ form a Pauli algebra acting on a qubit localized around site $L$, and  similarly for the operators $\eta^\oo_\mathrm{R}$ and $\D_\mathrm{R}^{(1)}$ at the right interface. 
These operators commute with the interface Hamiltonian and form a complete basis of operators acting on its ground space. These edge modes are protected by the projective algebra \eqref{proj.alg} and cannot be lifted by symmetric deformations of the Hamiltonian near the interfaces.

In general, the three SPT phases are distinguished by the projective algebra involving the local factors of $\D$ and $\eta^\ee, \eta^\oo$. The projective algebra implies that the \emph{defect} of the non-invertible symmetry $\D$ carries a non-trivial charge under $\eta^\ee, \eta^\oo$, or $\eta^\mathrm{d}$, which will be explored elsewhere. In the continuum, these charges are denoted by $\nu(\ee), \nu(\oo), \nu(\mathrm{d})$ in \cite{tambara2000representations,Thorngren:2019iar,Inamura:2021wuo}. Their values for the three SPT phases are\footnote{These invariants explain our terminology of `odd' and `even' SPT phases and suggest that the cluster state can be called the `diagonal' SPT. \label{terminology}}
\ie
	\nu_\text{cluster}(\ee) &= +1, &\nu_\text{cluster}(\oo) &= +1, &\nu_\text{cluster}(\mathrm{d}) &= -1 \,,\\
	\nu_\text{odd}(\ee) &= +1, &\nu_\text{odd}(\oo) &= -1, &\nu_\text{odd}(\mathrm{d}) &= +1 \,,\\
	\nu_\text{even}(\ee) &= -1,  &\nu_\text{even}(\oo) &= +1, &\nu_\text{even}(\mathrm{d}) &= +1 \,.
\fe
The projective signs \eqref{proj.alg} capture the relative charges  between $\ket{\text{cluster}}$ and $\ket{\text{odd}}$, corresponding to $\nu_\text{cluster}(\oo)\nu_\text{odd}(\oo) = -1$.

Similarly,  the interface between $\ket{\text{cluster}}$ and $\ket{\text{even}}$ is obtained by conjugating the previous interface system by the lattice translation $T$. For this interface, we find that the local factors of $\D$ are charged under $\eta^\ee$, leading to a projective algebra involving their local factors. This projective algebra is captured by $\nu_\text{cluster}(\ee)\nu_\text{even}(\ee) = -1$. 
We leave the interface between $\ket{\text{even}}$ and $\ket{\text{odd}}$ for the future.

\section{No symmetric entangler for non-invertible SPT phases \label{sec:entangler}}

\textbf{Relative classification of ordinary SPT phases:}
For  ordinary invertible global symmetries, without knowing the microscopic details,  it is only meaningful to discuss the relative SPT phase between two systems.
In continuum field theory, this corresponds to the ambiguity of adding a counterterm at short distances. 

In lattice systems with on-site symmetries, one often \emph{declares} that the product state is the trivial SPT state. This is because, in the microscopic spin variables, the product state has trivial entanglement and a non-trivial SPT state has a short-range entanglement that cannot be destroyed without breaking the symmetry. However, from the \emph{macroscopic} point of view, the notion of trivial SPT is only a choice and can be changed by doing a change of basis in the microscopic variables. 

Indeed,  there is usually a (locality-preserving) unitary operator $\mathsf{U}$ that commutes with the symmetry operators and maps the product state to any other SPT state.
This operator $\mathsf{U}$ is called the \emph{entangler}, since it maps the product state to a short-range entangled state.\footnote{A symmetric entangler might not always be realizable on a tensor-factorized Hilbert space. Its existence is related to the bulk-boundary correspondence between SPT states and anomalous symmetries on the boundary \cite[Appendix A]{Else:2014vma}.} A non-trivial entangler $\mathsf{U}$ must have a mixed 't Hooft anomaly with the original symmetry. 
The simplest example is the $ \bZ_2^\ee \times \bZ_2^\oo$ symmetric cluster entangler $\V$ in  \eqref{cluster.state}, which has a type III anomaly with $\bZ_2^\ee \times \bZ_2^\oo$.\footnote{While $\V$ is $\bZ_2^\ee \times \bZ_2^\oo$ symmetric, in the sense that it commutes with $\eta^\ee$ and $\eta^\oo$, the individual gates, $\mathrm{CZ}_{j,j+1}$, are not.  }

\textbf{Stacking of ordinary  SPT phases:}  
Given two $G$-SPT states for an invertible, finite symmetry $G$, their tensor product state is another SPT state with respect to the diagonal subgroup of $G\times G$. Hence, ordinary $G$-SPT phases can be multiplied via stacking. 
For instance, the 1+1d bosonic SPT phases  form a \textit{torsor} over the abelian group $H^2(G,U(1))$, which means that there is no canonical trivial element (identity).\footnote{This  is analogous to the space of spin structures of an orientable Riemannian manifold $M$, which is a torsor over $H^1(M,\bZ_2)$. This means that while there is no canonical trivial spin structure, the difference between any two spin structures is an element of $H^1(M,\bZ_2)$, i.e., a $\bZ_2$ connection. Similarly, while SPT phases have a relative classification, the differences between the SPT phases, i.e., the stacking action implemented by the entanglers, have an absolute classification. }  
Microscopically, the stacking is implemented by the symmetric entangler.
 
\textbf{No stacking of non-invertible SPT phases:}
However, for non-invertible symmetries, there is no notion of stacking operation for the following reason \cite{Thorngren:2019iar}. 
Starting with two systems with the same non-invertible symmetry, the tensor product system does not have a ``diagonal" non-invertible symmetry of the same type \cite{Bhardwaj:2017xup}. For instance, taking two systems with symmetries generated by $\eta^\ee,\eta^\oo, \D$, the ``diagonal" symmetry in the tensor product theory contains $\D \otimes \D$,  which obeys a different algebra since $(\D \otimes \D)^2 \neq (1 + \eta^\ee \otimes  \eta^\ee)(1 + \eta^\oo \otimes  \eta^\oo)$. 
In fact, the ``diagonal" symmetry is generally not closed under fusion, i.e., it is not a subcategory. 
Thus, there is no notion of invertibility for SPT phases protected by non-invertible symmetries.\footnote{Therefore, the adjective ``non-invertible" in ``non-invertible symmetry protected topological phase" is attached to the word ``symmetry", rather than to the word ``phase." (This is similar to the use of the term ``subsystem symmetry protected topological phase.")} 

In summary, stacking exists for invertible symmetries, so the notion of a trivial (invertible) SPT phase exists, but not in a canonical way. Indeed, the classification is relative, and declaring the product state as trivial is a \emph{choice}. Stacking does not exist for non-invertible symmetries, so the notion of a trivial non-invertible SPT phase does not exist, even if we specify a microscopic description.

\textbf{No symmetric entangler between non-invertible SPT phases:}
This suggests that there is no Rep(D$_8$)-symmetric entangler between the three different SPT states $\ket{\text{cluster}}, \ket{\text{odd}}, \ket{\text{even}}$.

We first prove a weaker statement that there cannot be a Rep(D$_8$)-symmetric \textit{Clifford} operator $\mathsf{U}$ such that, say,  $\mathsf{U} \ket{\text{cluster}}=\ket{\text{odd}}$. 
By definition, a Clifford unitary operator  maps the Pauli group to itself. Therefore, if there were such a Clifford operator,  it must map between the two stabilizer groups:  
\ie
&\mathsf{U}\langle Z_{j-1}X_j Z_{j+1}\rangle \mathsf{U}^{-1}\\
&  =\langle -Z_{2n-1} X_{2n}Z_{2n+1}, Y_{2n} X_{2n+1}Y_{2n+2}\rangle\,. 
\fe
This in particular implies that $Y_{2n}X_{2n+1}Y_{2n+2}$ can be written as a product of $\mathsf{U}Z_{j-1}X_j Z_{j+1} \mathsf{U}^{-1}$. However, this is a contradiction since the former does not commute with $\D$  (as discussed around \eqref{oddstabilizer}), while the latter does. 

We now argue for the more general statement without assuming $\mathsf{U}$ is Clifford. The idea is to show that if such a (unitary) symmetric entangler $\mathsf{U}$ exists, then there must exist a unitary operator $\hat{\mathsf{U}}$ in the $\mathbf{TST}$ gauged/KT transformed system such that it is invariant under the dual symmetry $\hat{\bZ}_2^\ee \times \hat{\bZ}_2^\oo \times \hat{\bZ}_2^\V$ and maps the two symmetry breaking phases to each other. However, this is impossible since the two SPT phases correspond to two different patterns of symmetry breaking after gauging and hence cannot be related by a symmetric unitary operator.

In other words, we claim that the existence of an entangler $\mathsf{U}$ satisfying
\ie
	\mathsf{U} \D &= \D \mathsf{U} \,, & ~~~ \mathsf{U} \eta^\ee &=  \eta^\ee \mathsf{U} \,, \\
	\mathsf{U} \ket{\text{cluster}} &= \ket{\text{odd}} \,, &\mathsf{U} \eta^\oo &=  \eta^\oo \mathsf{U} \,, \label{before.gauging}  
\fe
implies the existence of $\hat{\mathsf{U}}$ satisfying
\ie
	\hat{\mathsf{U}} \hat\V &= \hat\V \hat{\mathsf{U}} \,, & ~~~ \hat{\mathsf{U}} \hat{\eta}^\ee &=  \hat\eta^\ee \hat{\mathsf{U}} \,, \\
	\hat{\mathsf{U}} \ket{\widehat{\text{cluster}}} &= \ket{\widehat{\text{odd}}} \,, &\hat{\mathsf{U}} \hat\eta^\oo &=  \hat\eta^\oo \hat{\mathsf{U}} \,. \label{after.gauging} 
\fe
Here, $\ket{\widehat{\text{cluster}}}$ and $\ket{\widehat{\text{odd}}}$ are the symmetric ground states of $\hat H_\text{cluster}$ and $\hat H_\text{odd}$ in \emph{finite volume}. 
In infinite volume, each of these two states splits into 4 superselection sectors, which break the $\hat{\bZ}_2^\ee \times \hat{\bZ}_2^\oo$ symmetry spontaneously. 
We prove this claim  in Appendix \ref{app:no.entangler}  by showing that \eqref{after.gauging} is the image of \eqref{before.gauging} under the KT transformation, which maps $\mathsf{U}, \D, \ket{\text{cluster}}, \ket{\text{odd}}$ to $\hat{\mathsf{U}}, \hat\V, \ket{\widehat{\text{cluster}}}, \ket{\widehat{\text{odd}}}$, respectively.\footnote{Recall that KT transformation is implemented by a non-invertible operator that acts invertibly on the ${\bZ}_2^\ee \times {\bZ}_2^\oo$ invariant operators/states.}
However, \eqref{after.gauging} cannot be true since $\ket{\widehat{\text{cluster}}}$ represents a $\hat\bZ_2^\V$-preserving phase, while $\ket{\widehat{\text{odd}}}$ represents a spontaneous $\hat\bZ_2^\V$-breaking phase (see Section \ref{sec:3.SPTs}).

\section{Conclusions and outlook}

In the recent development of generalized symmetries, sometimes new symmetries are found in known quantum systems, and some other times they point to new systems.

In this paper we see an interesting interplay of these two scenarios. 
A new non-invertible symmetry $\D$ is discovered in the standard cluster model $H_\text{cluster}$, which, together with the $\bZ_2^\ee \times\bZ_2^\oo$ symmetry,  forms a Rep(D$_8$) fusion category. 
This  symmetry  leads to two new commuting Pauli Hamiltonians $H_\text{odd}$  and $H_\text{even}$, whose ground states $\ket{\text{odd}}$ and $\ket{\text{even}}$ are SPT states protected by Rep(D$_8$). 
They are distinguished from the cluster Hamiltonian by their symmetry breaking patterns after the KT transformation/$\mathbf{TST}$ gauging.

While $\ket{\text{cluster}}, \ket{\text{odd}},\ket{\text{even}}$ are in the same SPT phase as far as the $\bZ_2^\ee\times \bZ_2^\oo$ symmetry is concerned,  they are distinguished by the non-invertible symmetry. 
This is reflected in the edge modes from the local projective algebra of the non-invertible symmetry.
In contrast, the product state is not invariant under $\D$, and there is no canonical notion of a trivial non-invertible SPT phase.

Finally, we argue that there is no Rep(D$_8$) symmetric entangler relating these three SPT states. 
This observation is related to the fact that there is no stacking operation for systems with non-invertible symmetries. 
Moreover, it suggests that the classification of non-invertible SPT phases is not relative. 
This means that two distinct non-invertible SPT phases correspond to two different continuum field theories at long distances that do not differ just by    a local counterterm. 
This is unlike the case of ordinary invertible symmetries, where the individual SPT phases have the same behavior at long distances, and their distinction is only visible from the edge modes on the interface between two such phases.

Our lattice construction  provides some of the simplest examples of non-invertible SPT phases. 
First, our Hilbert space is a  tensor product  of qubits, and the Hamiltonians are written in terms of the Pauli operators.  
Second,  the Rep(D$_8$) fusion category is one of the simplest (non-anomalous) non-invertible symmetries that admit  SPT phases, and we constructed the lattice models for \textit{all} of its possible SPT phases.

Our models are complementary to the non-invertible SPT models of \cite{Inamura:2021szw,Fechisin:2023dkj}, which employ more general local Hilbert spaces than qubits. 
In particular, our Rep(D$_8$) symmetry of the standard cluster model is different from the $G\times \text{Rep($G$)}$ symmetry of   the generalized cluster model discussed in \cite{Fechisin:2023dkj} . 
When $G=\bZ_2$, their model reduces to the standard cluster model \eqref{cluster.H}, but their category symmetry reduces to the ordinary  $\bZ_2 \times \text{Rep}(\bZ_2)\simeq \bZ_2^\ee\times \bZ_2^\oo$ symmetry, which is an invertible subgroup of our Rep(D$_8$) fusion category.

There are several future directions:
\begin{itemize}
\item The key to our construction is that many of the non-invertible symmetries become (anomalous) invertible symmetries after gauging. 
Thus, the different non-invertible SPT phases are distinguished by the conventional spontaneous symmetry breaking patterns in the gauged systems. This construction can be readily generalized to many other non-invertible symmetries.
\item It would be interesting to explore the anomaly inflow for non-invertible symmetry by the explicit non-invertible SPT lattice models.
\item The Rep(D$_8$) fusion category, along with its two other cousins Rep(Q$_8$) and Rep(H$_8$),  are some of the simplest non-invertible symmetries that can be gauged. In Appendix \ref{app:D8} we gauge Rep(D$_8$) in steps, but it would be desirable to directly gauge the entire category on the lattice, mirroring the results of \cite{Choi:2023vgk,Diatlyk:2023fwf,Perez-Lona:2023djo}. 
\item Generalization to higher dimensions.
\end{itemize}

\begin{acknowledgments}

We thank Nati Seiberg for discussions which initiated this project. We are grateful to  Yichul Choi, Abhinav Prem, Wilbur Shirley, Nikita Sopenko, Nathanan Tantivasadakarn, Ruben Verresen, Tzu-Chieh Wei, Carolyn Zhang, and Yunqin Zheng for interesting discussions. 
We thank Da-Chuan Lu, Nikita Sopenko, and Yifan Wang for comments on the draft. 
SS gratefully acknowledges support from the U.S. Department of Energy grant DE-SC0009988, the Sivian Fund, and the Paul Dirac Fund at the Institute for Advanced Study. 
SHS is  supported by the Simons Collaboration on Ultra-Quantum Matter, which is a grant from the Simons Foundation (651444, SHS). 
 The authors of this paper were ordered alphabetically.

\end{acknowledgments}

\appendix

\section{Lattice gauging}

In this appendix we review some standard facts about  the untwisted ($\mathbf{S}$) and twisted gauging ($\mathbf{TST}$)  of a $\bZ_2^\ee \times \bZ_2^\oo$ symmetry \eqref{Z2Z2} 
on the lattice. 
These two gaugings are also known as (the square of) the Kramers-Wannier transformation and the KT transformation, respectively. 
Our discussion follows Appendix A of \cite{Seiberg:2024gek} closely.

We work with a closed chain of $L$ sites, with the Pauli operators $X_j,Z_j$ acting on each site.  
We assume a general finite-range Hamiltonian $H$ that is $\bZ_2^\ee \times \bZ_2^\oo$ symmetric (such as the trivial Hamiltonian $H_\text{trivial}$ or the cluster Hamiltonian $H_\text{cluster}$).

 \subsection{KW$^{\otimes2}$:~untwisted lattice gauging $(\mathbf{S})$}\label{app:S}

To gauge the $\bZ_2^\ee $ symmetry, we introduce a gauge field qubit on every odd site. We denote the corresponding Pauli operators by $\sigma^x_{2n-1},\sigma^z_{2n-1}$. 
Similarly, to gauge the $\bZ_2^\oo$ symmetry, we introduce a gauge field qubit on every even site, and denote the corresponding Pauli operators by $\sigma^x_{2n} ,\sigma^z_{2n}$. We impose the $\bZ_2^\ee \times\bZ_2^\oo$ Gauss law at every site:
\ie\label{gauss}
G_{j} = \sigma^z_{j-1}   X_{j} \sigma^z_{j+1} = 1\,.
\fe
We then couple the gauge fields to the original Hamiltonian $H$ minimally by replacing every $Z_{j-1} Z_{j+1}$ by $Z_{j-1} \sigma^x_j Z_{j+1}$. 
This leads  to a gauged Hamiltonian $H'$, written in terms of $X_j,Z_j, \sigma^x_j, \sigma^z_j$.

Not every   Pauli operator is gauge invariant and commutes with the Gauss law constraint. 
To proceed, we define a new set of gauge-invariant variables $X_j',Z_j'$ that commute with the Gauss law:
\ie
X_j' =   Z_{j-1} \sigma^x_j Z_{j+1}\,,~~~Z'_j  =\sigma^z_j \,.
\fe
Clearly, $X_j',Z_j'$ obey the standard Pauli algebra.

Now we can write the gauged Hamiltonian in terms of new, gauge-invariant variables. The minimal coupling term becomes $Z_{j-1}\sigma^x_jZ_{j+1}= X_j'$, and the operator $X_j$ is replaced by $Z_{j-1}'Z_{j+1}'$ using the Gauss law. All in all, the untwisted gauging maps the  $\bZ_2^\ee \times \bZ_2^\oo$ local operators as
\ie\label{Smap}
\mathbf{S}:~~~X_j \rightsquigarrow Z_{j-1} ' Z_{j+1}' \,,~~~~
Z_{j-1} Z_{j+1} \rightsquigarrow X_j'\,.
\fe
We use $\rightsquigarrow$ to emphasize that this transformation is not implemented by  a unitary operator. 
Indeed, if we identify the new variables with the original ones $X_j' = X_j ,Z_j'=Z_j$, then \eqref{Smap} is implemented by the non-invertible operator $\D$ as in  \eqref{Daction}:
\ie\label{appDaction}
\D X_j  =Z_{j-1} Z_{j+1} \D\,,~~~
\D Z_{j-1}Z_{j+1} = X_j \D\,.
\fe

In the gauged Hamiltonian $H'$,  the original $\bZ_2^\ee \times \bZ_2^\oo$ global symmetry $\eta^\ee,\eta^\oo$ in \eqref{Z2Z2} is gauged and is no longer a global symmetry of $H'$. Rather, the gauged Hamiltonian is invariant under a dual global $\bZ_2^{\ee'} \times \bZ_2^{\oo '}$ symmetry  generated by
\ie
\eta^{\ee ' } = \prod_{j:\text{even}} X_j' \,,~~~
\eta^{\oo ' } = \prod_{j:\text{odd}} X_j' \,.
\fe
This is the lattice version of the dual symmetry (also known as the quantum symmetry or the orbifold symmetry) in \cite{Vafa:1989ih}.

Applying the untwisted gauging on the trivial and the cluster Hamiltonians
\ie\label{trivialcluster}
H_\text{trivial} &= - \sum_{j=1}^L X_j\,,\\
H_\text{cluster} &=- \sum_{j=1}^L Z_{j-1}X_j Z_{j+1}\,,
\fe
we find the two gauged Hamiltonians in \eqref{S}
\ie\label{app.S}
	H'_\text{trivial}  &= -  \sum_{j=1}^L\, Z_{j-1}' Z_{j+1}' \,,\\
	H'_\text{cluster}  &= -  \sum_{j=1}^L \, Z_{j-1}' X_j'  Z_{j+1}' \,.
\fe
We see that the trivial phase becomes a $\bZ_2^{\ee' }\times \bZ_2^{\oo'}$ spontaneously symmetry broken phase, while the cluster phase remains invariant and preserves the global symmetry.

Via gauging, one can distinguish the two SPT phases by the string order parameters. 
In the two gauged models, the local order parameters are $\langle Z_{2n-1} '\rangle$ and $\langle Z_{2n} '\rangle$, which distinguish the spontaneous symmetry breaking phase from the unbroken phase. For large $l$, $\langle Z_{0}' Z_{2l}' \rangle \sim \langle Z_{0}' \rangle \langle Z_{2l}' \rangle$ is equal to $0$ and $1$ in the symmetry preserving and breaking phases, respectively, up to exponentially small corrections. 
Before gauging, these local, order parameters come from certain non-local, string-like order parameters. 
More specifically, doing an inverse of \eqref{KW2}, we find the string order parameter $\langle X_j X_{j+2} \cdots X_{j+2l} \rangle$ that distinguishes the trivial state from the cluster state. Their correlation functions are 
\ie
	\langle X_j X_{j+2} \cdots X_{j+2l} \rangle \sim \begin{cases} \text{constant} & \text{product state} \\
	0 & \text{cluster state}
	\end{cases} \label{string.order}
\fe
up to corrections exponentially small in $l$.

\subsection{KT:~twisted  lattice gauging $(\mathbf{TST})$}\label{app:TST}

The twisted gauging $\mathbf{TST}$ is related to the untwisted one  $\mathbf{S}$ by conjugating with the cluster entangler $\V=\prod_j \mathrm{CZ}_{j,j+1}$, which implements the $\mathbf{T}$ transformation:
\ie
\mathbf{T}: ~~ \begin{array}{l} X_j \mapsto \V X_j\V^{-1} = Z_{j-1}X_j Z_{j+1}\,,\\
Z_j \mapsto \V Z_j\V^{-1} = Z_j\,. \end{array}
\fe

Let us break down the $\mathbf{TST}$ gauging in steps. 
 First, we conjugate the Hamiltonian by $\V$. 
 Second, we  perform the untwisted gauging to map it to a gauged Hamiltonian as in Appendix \ref{app:S}. 
We denote the gauge-invariant variables in the gauged Hamiltonian as $\hat X_j,\hat Z_j$, to distinguish from the primed variables for the untwisted gauging. 
Third, we  conjugate the gauged Hamiltonian by 
\ie\label{hatV}
\hat \V =  \prod_{j=1}^L \hat{\mathrm{CZ}}_{j,j+1} =\prod_{j=1}^L {1+\hat Z_j +\hat Z_{j+1}-\hat Z_j \hat Z_{j+1}\over 2}
\fe
This sequence of operation results in the following gauging map, commonly denoted as $\mathbf{TST}$ in the field theory literature, on the $\bZ_2^\ee \times \bZ_2^\oo$ symmetric local operators:
\ie\label{appTST}
\mathbf{TST}:~~~ X_j \rightsquigarrow \hat X_j \,,~~~
Z_{j-1}Z_{j+1} \rightsquigarrow \hat Z_{j-1} \hat X_j \hat Z_{j+1}\,.
\fe
Again, we use $\rightsquigarrow$ to emphasize that this transformation is not implemented by  a unitary operator. 
Indeed, if we identify the new variables with the original ones $\hat X_j= X_j ,\hat Z_j=Z_j$, then \eqref{Smap} is implemented by the non-invertible operator $\V\D \V$ \cite{ParayilMana:2024txy}:
\ie\label{VDV}
&(\V\D \V) X_j = X_j (\V\D \V)\,,\\
&(\V\D \V) Z_{j-1}Z_{j+1} = Z_{j-1}X_j Z_{j+1} (\V\D \V)\,.
\fe
This is the KT transformation.

The twisted gauging maps the original Hamiltonian $H$ of $X_j,Z_j$ to 
a gauged Hamiltonian $\hat H$ of $\hat X_j, \hat Z_j$. Again, the gauged Hamiltonian  is invariant under a dual global $\hat\bZ_2^{\ee} \times \hat\bZ_2^{\oo }$ symmetry  generated by
\ie\label{hateta}
\hat\eta^{\ee  } = \prod_{j:\text{even}} \hat X_j\,,~~~
\hat \eta^{\oo } = \prod_{j:\text{odd}} \hat X_j \,.
\fe
As examples, applying the twisted gauging on $H_\text{trivial}$ and $H_\text{cluster}$  gives \eqref{TST}.

We summarize these operations in Table \ref{tab:gauging}.

\begin{table}[]
\begin{tabular}{|c|c|c|}
\hline
operation & transformation & operator   \\ 
\hline
$\mathbf{S}$ gauging  & $ X_j \rightsquigarrow Z'_{j-1} Z'_{j+1} $ & $\D$ \\
 ($\text{Kramers-Wannier}^{\otimes 2}$)  & $ Z_{j-1} Z_{j+1} \rightsquigarrow X'_j $ &  \\ \hline
$\mathbf{TST}$ gauging &$X_j \rightsquigarrow \hat{X}_{j}  $ &$\V \D \V$\\
 (Kennedy-Tasaki) 
& $Z_{j-1} Z_{j+1} \rightsquigarrow  \hat{Z}_{j-1} \hat{X}_j \hat{Z}_{j+1}$ &  \\ \hline
 $\mathbf{T}$:~stacking with SPT & $X_j \mapsto Z_{j-1} X_j Z_{j+1}$ & $\V$
 \\ (Cluster entangler)& $Z_{j} \mapsto Z_j $ & \\ \hline
\end{tabular}
\caption{The table discusses two different gaugings of a $\bZ_2\times \bZ_2$ global symmetry, which in the continuum field theory literature are often denoted as $\mathbf{S}$ and $\mathbf{TST}$, as well as the stacking by the cluster SPT phase $(\mathbf{T})$. The middle column shows the action of these transformations on the $\bZ_2\times\bZ_2$ symmetric local operators. The rightmost column gives the lattice operators implementing these transformations if we identify the gauged and ungauged variables by dropping the primes/hats.}
\label{tab:gauging}
\end{table}

 \section{Non-invertible symmetry under gauging}\label{app:t.gauging}

So far we have only assumed the original Hamiltonian  $H$ is invariant under an invertible $\bZ_2^\ee \times \bZ_2^\oo$ global symmetry.   
The twisted gauged Hamiltonian $\hat H$ is then invariant under a dual $\hat\bZ_2^\ee \times \hat\bZ_2^\oo$ symmetry. 
Now we further assume $H$ commutes with the non-invertible global symmetry $\D$. 
What happens to $\D$ under the twisted gauging $\mathbf{TST}$?

\subsection{Rep(D$_8$) $\rightarrow \hat\bZ_2^\ee \times \hat \bZ_2^\oo \times \hat \bZ_2^\V$ \label{D.after.KT}}
 
Below we argue that the non-invertible symmetry $\D$ of $H$ becomes an invertible $\hat\bZ_2^\V$ symmetry, generated by $\hat \V$ in \eqref{hatV}    after the twisted gauging.   
More precisely, we will show that any Hamiltonian that is invariant under $\eta^\ee, \eta^\oo , \D$ (which forms a Rep(D$_8$) fusion category) is mapped to a gauged Hamiltonian $\hat H$ that is invariant under a dual $\hat \bZ_2^\ee \times \hat\bZ_2^\oo \times\hat \bZ_2^\V
$ symmetry generated by $\hat \eta^\ee,\hat \eta^\oo ,\hat \V$ under the twisted gauging. 

 For instance, the double-Ising Hamiltonian
\begin{align}\label{Ising2}
 H_{\text{Ising}^2} =- \sum_{j=1}^{L}  \left( Z_{j-1}  Z_{j+1} +  X_j  \right)
\end{align}
is invariant under $\D,\eta^\ee,\eta^\oo$. Applying the $\mathbf{TST}$ gauging turns it into 
\ie\label{LG}
 \hat H_\text{LG} = -  \sum_{j=1}^{L} \left(\hat Z_{j-1}\hat  X_j \hat Z_{j+1} +\hat X_j  \right)  \,,  
\fe
which indeed is invariant under $\hat\bZ_2^\ee \times \hat\bZ_2^\oo \times \hat\bZ_2^\V$. 
This Hamiltonian (up to a unitary transformation) is   the boundary theory for the 2+1d Levin-Gu SPT phase labeled by $H^3(\mathbb{Z}_2,U(1))=\mathbb{Z}_2$ \cite{Levin:2012yb}.

To prove this generally, it suffices to focus on the action of $\D$ on the local terms in the Hamiltonian $H$, which are all  $\bZ_2^\ee \times \bZ_2^\oo$ symmetric. 
On the $\bZ_2^\ee \times \bZ_2^\oo$ symmetric sector, the KT transformation \eqref{appTST} is invertible, and we can  conjugate  the action of $\D$ in \eqref{appDaction} by the KT transformation.  
This results in a transformation in the gauged model that acts in the same way as $\hat \V$:
\ie
\hat \V:~~ \begin{array}{l} \hat X_j \mapsto \hat Z_{j-1} \hat X_j \hat Z_{j+1}\,,\\
\hat Z_{j-1} \hat Z_{j+1}\mapsto \hat Z_{j-1} \hat Z_{j+1}\,. \end{array}
\fe
This shows at the level of the Hamiltonians that $\D$ becomes $\hat\V$ after the twisted gauging.

The fact that $\D$ becomes $\hat \V$ can be understood more intuitively if we identify the hatted variables in the gauged Hamiltonian with the original one as $\hat{X}_ j = X_j$, $\hat{Z}_ j = Z_j$ and $\hat \V = \V$. Under this identification, the twisted gauging/KT transformation is implemented by $\V \D \V$ in the sense of \eqref{VDV}. 
More generally, given any $\bZ_2^\ee \times \bZ_2^\oo$ symmetric operator $O=O(X_j,Z_j)$, its image under the KT transformation is given by $\hat{O}=\hat O(X_j,Z_j)$ satisfying
\ie
	(\V \D \V) \, O = \hat{O} \, ( \V \D \V)\,. \label{noninv.action}
\fe
Now the fact that  $\D$ becomes $\V$  under the KT transformation follows from the following identity
\ie
	(\V\D \V) \D  = 2\V (\V\D \V) \,. \label{VDVD}
\fe
This relation is reminiscent of the  relation $\mathbf{TSTS} = \mathbf{ST}$ in the modular group.

To show this relation, we first note that both the left and right hand sides of the equation above have the same action, in the sense of \eqref{noninv.action}, on $\bZ_2^\ee \times \bZ_2^\oo$ symmetric operators. This implies that the restriction of these two operators to the $\bZ_2^\ee \times \bZ_2^\oo$ symmetric sector is the same up to an overall constant, i.e., 
\ie
	\Pi \, (\V\D \V\D )\, \Pi  = c \, \Pi \, (2 \D \V) \, \Pi \,,
\fe
where $\Pi = \frac{1+\eta^\ee}{2}\frac{1+\eta^\oo}{2}$
and $c \in \mathbb{C}$. Since $\Pi$ commutes with $\V$ and $\Pi \D = \D \Pi = \D$, we find that $\V\D \V\D  = c \, 2 \D \V$.

To fix the overall constant we compare the action of these operators on the cluster state. We begin by 
\ie
	 \D \V \ket{\text{cluster}} &=  \D \ket{++ \cdots +} \\
	 &= \D^\ee \D^\oo T \ket{++ \cdots +} \\
	 &= \ket{00} + \ket{01} + \ket{10} + \ket{11} \,, \label{DV.cluster}
\fe
where $\ket{z_\ee z_\oo} = \ket{z_\ee z_\ee \cdots z_\ee }_\text{even} \otimes \ket{z_\oo z_\oo \cdots z_\oo}_\text{odd}$ for $z_\ee, z_\oo = 0,1$. This follows from \cite{Seiberg:2024gek}
\ie
	\D^\ee \ket{++ \cdots + }_\text{even} &= \ket{00 \cdots 0 }_\text{even} + \ket{11 \cdots 1 }_\text{even}\,, \\
	\D^\oo \ket{++ \cdots + }_\text{odd} &= \ket{00 \cdots 0 }_\text{odd} + \ket{11 \cdots 1 }_\text{odd} \,.
\fe
Next, we consider the action of $\V\D \V\D$:
\ie
	\V\D \V\D \ket{\text{cluster}} &= 2 \V\D \ket{++ \cdots +}  \\
	&= 2 \V \big(\ket{00} + \ket{01} + \ket{10} + \ket{11} \big) \\
	&= 2 \big(\ket{00} + \ket{01} + \ket{10} + (-1)^L \ket{11} \big) \,. \label{VDVD.cluster}
\fe
Here we have used the relation $\D \ket{\text{cluster}} = 2\ket{\text{cluster}}$, proven in Appendix \ref{app:D.action}, and the fact that
\ie
	\mathrm{CZ}_{j,j+1} \ket{z_\ee z_\oo} = (-1)^{z_\ee z_\oo} \ket{z_\ee z_\oo} \,.
\fe
Since $L$ is taken to be even throughout, comparing \eqref{DV.cluster} with \eqref{VDVD.cluster} fixes  $c=1$.  We have thus proved   \eqref{VDVD}.

\subsection{Type III anomaly and the Levin-Gu model}\label{app:typeIII}

We now show that the dual 
$\hat \bZ_2^\ee \times \hat\bZ_2^\oo \times\hat \bZ_2^\V
$ symmetry  generated by \eqref{hatV} and \eqref{hateta} has an 't Hooft anomaly of type III.

While the anomaly of the symmetry operators $\hat\eta^\ee,\hat\eta^\oo, \hat \V$ does not depend on the choice of the symmetric Hamiltonian, for concreteness, we consider the Levin-Gu Hamiltonian in \eqref{LG}. 
(We assume $L$ to be a multiple of 4 for simplicity.)

It is known that  the diagonal $\bZ_2$ symmetry generated by 
\ie
\hat\eta^\ee \hat \eta^\oo \hat \V = \prod_{j=1}^L \hat X_j  \,\hat{\mathrm{CZ}}_{j,j+1}\,.
\fe
has  an 't Hooft anomaly described by $e^{ i\pi \int \hat A\cup \hat A\cup \hat A }$ \cite{Chen:2011bcp} (see also \cite{Wang:2017loc}). 
Below we  show that the full $\hat\bZ_2^\ee\times \hat \bZ_2^\oo \times \hat \bZ_2^\V$ symmetry    has an 't Hooft anomaly described by the 2+1d invertible field theory
\ie
e^{i \pi \int \hat A^\ee \cup\hat A^\oo \cup \hat A^\V}\,,
\fe
where $\hat A^\ee,\hat A^\oo,\hat A^\V$ are  background $\bZ_2$ gauge fields. 
This is referred to as a type III anomaly  \cite{deWildPropitius:1995cf,Wang:2014tia} because it involves all three generators of $\hat\bZ_2^\ee\times \hat \bZ_2^\oo\times \hat\bZ_2^ \V$. 
As a consistency check, this reduces to the Levin-Gu anomaly for the diagonal symmetry by setting $\hat A^\ee =\hat A^\oo = \hat A^\V$. 
One feature of this type III anomaly is that out of the 7 $\bZ_2$ subgroups of $\hat\bZ_2^\ee\times \hat \bZ_2^\oo \times \hat \bZ_2^\V$, only the one generated by $\hat\eta^\ee \hat \eta^\oo \hat \V$ has an anomaly.

One diagnostic of the  type III anomaly is the projective algebra of two of the $\bZ_2$ operators in the Hamiltonian twisted by the third $\bZ_2$. 
See  \cite{Cheng:2022sgb,Seifnashri:2023dpa} (and also \cite{Yao:2020xcm}) for general discussions of anomalies on the lattice. 
Consider the Levin-Gu Hamiltonian twisted by the $\hat\eta^\text{e}$ symmetry at site $1$:
\ie
\hat H_{\eta^\text{e}}  =& \cdots 
- \left(\hat X_{L}+\hat Z_{L-1} \hat X_{L}\hat Z_1  \right)\\
&
- \left(\hat X_{1} - \hat Z_{L} \hat X_{1} \hat Z_2  \right)
- \left(\hat X_{2}+\hat Z_{1}\hat X_{2} \hat Z_3  \right)
-\cdots \,,
\fe
where $\cdots$ represents terms of the form $\hat Z_{j-1}\hat X_j \hat Z_{j+1}+\hat X_j$. 
The symmetry operators in the $\hat\eta^\text{e}$-twisted problem are
\ie
\hat\eta^\text{e}_{\text{e}}  = \prod_{j:\text{even}}\hat X_j \,,~~~
\hat\eta^\text{o}_{\text{e}} = \prod_{j:\text{odd}}\hat X_j\,,~~~
\hat \V_{\text{e}} =  \hat Z_1\hat \V \,,
\fe
where the subscript e means that these are symmetry operators that commute with the $\hat \eta^\ee$-twisted Hamiltonian $\hat H_{\eta^\text{e}} $. 
We find a projective algebra in this twisted problem:
\ie
\hat \eta^\text{o}_{\text{e}}\hat  \V_\ee
=- \hat \V_\ee\hat \eta^\text{o}_\ee\,,
\fe
which is the hallmark of the type III anomaly. 
Related projective algebras have also been discussed in \cite{Wang:2014tia}.

To summarize, the twisted gauging maps a non-invertible symmetry Rep(D$_8$) to an anomalous invertible symmetry $\hat\bZ_2^\ee \times \hat \bZ_2^\oo \times\hat \bZ_2^V$ as in \eqref{D8typeIII}. 
This fact was known in the continuum 
\cite{Tachikawa:2017gyf,Kaidi:2023maf}, and here we give a lattice derivation. 
It is also consistent with the fact that the two fusion categories Rep(D$_8$) and Vec$_{\bZ_2^3}^\omega$ (with $\omega\in H^3(\bZ_2^3,U(1))$ chosen to be the type III cocycle \cite{deWildPropitius:1995cf}) have the same Drinfeld center, i.e., the same 2+1d topological order (see, e.g.,  \cite{Iqbal:2023wvm,Bhardwaj:2024qrf}).
 More generally, it is common that an anomalous invertible symmetry is mapped to a non-invertible symmetry under gauging, and vice versa \cite{Kaidi:2021xfk}. 
 For instance, the non-invertible Kramers-Wannier symmetry of the critical Ising model arises from the chiral fermion parity \cite{Ji:2019ugf,Lin:2019hks,Seiberg:2023cdc}, and the non-invertible chiral symmetry of the 3+1d quantum electrodynamics arises from the Adler-Bell-Jackiw anomaly \cite{Choi:2022jqy,Cordova:2022ieu}.

\subsection{Gauging Rep($\mathrm{D}_8$) in steps \label{app:D8}}

The algebra \eqref{ni.symm} implies that $\D, \eta^\ee,\eta^\oo$ form  one of the four $\mathbb{Z}_2^\ee \times \mathbb{Z}_2^\oo$ Tambara-Yamagami (TY) fusion categories \cite{TAMBARA1998692} (see also \cite{Bhardwaj:2017xup,Thorngren:2019iar,Perez-Lona:2023djo}). 
Among these four TY categories, three of them admit fiber functors (i.e, the mathematical description of the non-invertible SPT phases), which are Rep(D$_8$), Rep(Q$_8$), and Rep(H$_8$). 
Here Q$_8$ is the quaternion group of order 8 and H$_8$ is  a non-grouplike Hopf algebra of dimension 8 known as the Kac-Paljutkin algebra. 
These three fusion categories are free of generalized 't Hooft anomaly both in the sense that they can be gauged \cite{Choi:2023vgk,Diatlyk:2023fwf,Perez-Lona:2023djo}, and they are compatible with a trivially gapped phase, i.e., a non-invertible SPT phase.
More specifically, Rep(D$_8$) admits three fiber functors, while Rep(Q$_8$), and Rep(H$_8$) each has a unique fiber functor \cite{TAMBARA1998692,Thorngren:2019iar,Inamura:2021wuo}. 

Here we  show that our lattice symmetry $\D,\eta^\ee,\eta^\oo$ form the Rep(D$_8$). 
To prove this, we use the following fact. 
Gauging a finite $G$ global symmetry in   1+1d  gives rise to a dual global symmetry described by Rep($G$) \cite{Bhardwaj:2017xup} (see also \cite{Ji:2019jhk}).  
Below we gauge the non-invertible symmetry generated by $\D,\eta^\ee,\eta^\oo$ and show that the gauged theory has a dual  D$_8$ symmetry. 
This determines the fusion category of the ungauged system to be Rep(D$_8$).

Instead of gauging the non-invertible symmetry Rep(D$_8$) in one shot, we do it in two steps, with each step only consisting of gauging ordinary invertible symmetries. 
In the first step, we twisted gauge the $\bZ_2^\ee\times\bZ_2^\oo$ symmetry, i.e., the $\mathbf{TST}$ gauging, i.e., the KT transformation. 
This was done in Appendix \ref{app:t.gauging}, which turns   the non-invertible symmetry \eqref{ni.symm} into an anomalous $\hat\bZ_2^\ee \times \hat\bZ_2^\oo \times \hat\bZ_2^\V$ symmetry generated by $\hat\eta^\ee$, $\hat\eta^\oo$, and $\hat \V$.

In the second step, we gauge the $\hat\bZ_2^\V$ symmetry generated by $\hat \V$. To do this gauging, we introduce gauge fields $\tilde{X}_{j}$ on sites and impose the Gauss law constraints
\ie
G_{j,j+1} = \tilde{X}_j \hat{\mathrm{CZ}}_{j,j+1} \tilde{X}_{j+1}
\fe
to obtain the Hilbert space of the (second) gauged model. 
In this model, we find a dual D$_8$ symmetry generated by
\ie
	\tilde{\eta}^\ee &= \prod_{j: \text{even}} \hat X_j \prod_{j: \text{odd}} \frac{1+\hat Z_j + \tilde{Z}_j -\hat Z_j \tilde{Z}_j }{2} \,, \\
	\tilde{\eta}^\oo &= \prod_{j: \text{odd}} \hat X_j \prod_{j: \text{even}} \frac{1+\hat Z_j + \tilde{Z}_j -\hat Z_j \tilde{Z}_j }{2} \,, \\
	\tilde \V &= \prod_{j} \tilde{Z}_j \,. \label{D8}
\fe
One can check the operators above commute with the Gauss's law constraints and generate a D$_8$ symmetry since $(\tilde{\eta}^\ee)^2 = (\tilde{\eta}^\oo)^2 = \tilde \V^2 = 1$ and $\tilde{\eta}^\ee \tilde{\eta}^\oo= \tilde \V \, \hat{\eta}^\oo\hat{\eta}^\ee$.

As a concrete example, we again start with the double-Ising Hamiltonian in \eqref{Ising2} which has a Rep(D$_8$) symmetry. 
The first step of $\mathbf{TST}$ gauging turns it into the Levin-Gu model $\hat H_\text{LG}$ in \eqref{LG}. 
In the second step,  gauging $\hat\bZ_2^\V$ turns $\hat H_\text{LG}$ into
\ie
	&\tilde H = \\ &
	-  \sum_{j=1}^{L} \hat X_j \left( \frac{1+\tilde{Z}_j}{2} (1+ \hat Z_{j-1}\hat Z_{j+1}) + \frac{1-\tilde{Z}_j}{2} (\hat Z_{j-1} + \hat Z_{j+1})  \right) 
\fe
which indeed is invariant under the D$_8$ symmetry generated by \eqref{D8}.  

To conclude, we gauge the symmetry generated by $\eta^\ee ,\eta^\oo, \D$ of a general Hamiltonian in two steps:
\ie
&\text{Rep(D$_8$)}~ \xrightarrow[\mathbf{TST}]{\text{gauge }\mathbb{Z}_2^\ee\times\bZ_2^\oo }  ~\hat\bZ_2^\ee \times \hat\bZ_2^\oo \times \hat \bZ_2^\V~
 \xrightarrow{\text{gauge }\hat\bZ_2^\V}  ~\text{D$_8$}
\fe
In the end, we find an ordinary global symmetry D$_8$ in the  gauged Hamiltonian. This implies that the original non-invertible symmetry is Rep(D$_8$).

\section{SPT phases as invertible field theories}\label{app:cont}

In this appendix we review the continuum description of the $\bZ_2 \times \bZ_2$ SPT phases as invertible field theories (see, e.g., \cite{Freed:2014eja}). We then gauge these SPT phases in two different ways. 
Our discussion follows \cite{Thorngren:2019iar,Li:2023ani} closely. 

We first review the gauging of a non-anomalous $\bZ_2$ global symmetry in 1+1d quantum field theory (QFT).  
The generalization to gauging a $\bZ_2\times\bZ_2$ symmetry will be straightforward. 
Consider a general 1+1d QFT with a $\bZ_2$ global symmetry. Let the partition function  on a closed spacetime 2-manifold $\Sigma$ be $Z[A]$, where $A\in H^1 (\bZ_2,U(1))$ is the background $\bZ_2$ gauge field. 
The gauged QFT  has a dual symmetry $\bZ_2'$ \cite{Vafa:1989ih}. 
Let $A'$ be the background gauge field for $\bZ_2'$, then the partition function of the gauged theory is
\ie\label{SZ2}
Z' [A' ]  = {1\over \sqrt{|H^1 (\Sigma , \bZ_2) |}} \sum_{a\in H^1 (\Sigma, \bZ_2)} Z[a] \exp\left[ i\pi \oint_\Sigma a\cup A' \right] \,.
\fe
The normalization is up to the ambiguity of an Euler counterterm $\lambda^{\chi(\Sigma)}$ with $\lambda>0$.  
From \eqref{SZ2}, it is clear that if we further gauge the dual $\bZ_2'$ global symmetry of the gauged QFT,  we retrieve the original QFT.

The continuum description of a product state $\ket{++...+}$ can be chosen to be the trivial QFT, whose partition function is 1 on every closed spacetime 2-manifold $\Sigma$:
\ie
Z_\text{trivial} [A^\ee ,A^\oo]= 1\,.
\fe
Once this choice is made, the continuum description for the cluster state is given by a nontrivial invertible field theory of the $\bZ_2^\ee \times \bZ_2^\oo$ background gauge fields $A^\ee ,A^\oo$:
\ie\label{Zcluster}
Z_\text{cluster}[A^\ee ,A^\oo] = \exp \left[ i \pi \oint_M A^\ee \cup A^\oo\right]
\fe
Note that $A^\ee ,A^\oo$ are classical, background $\bZ_2$ gauge fields, and there is no path integral over them. Therefore, the partition function of this invertible field theory is always a phase on every closed 2-manifold $\Sigma$ \cite{Freed:2004yc}.

\subsection{$\mathbf{S}$ gauging in the continuum}

 We now gauge the $\bZ_2^\ee \times \bZ_2^\oo$ global symmetry  of these two invertible field theories. 
 This (untwisted) gauging is referred to as the $\mathbf{S}$ gauging in the literature (e.g., \cite{Witten:2003ya,Gaiotto:2014kfa,Bhardwaj:2020ymp,Li:2023ani}), representing the corresponding element in the modular group.
 
 We promote the background gauge fields to be dynamical and denote them by $a^\ee,a^\oo$. 
We also introduce the background gauge fields $A^{\ee '},A^{\oo '}$ for the dual $\bZ_2^{\ee '} \times\bZ_2^{\oo '}$ global symmetry. 
Gauging the trivial theory gives:
\ie
~&Z' _\text{trivial} [A^{\ee' },A^{\oo'}] = {1\over \sqrt{|H^1 (\Sigma , \bZ_2\times \bZ_2) | } }\\
&\times\sum_{(a^\ee, a^\oo) \in H^1 (\Sigma,\bZ_2\times\bZ_2)} \exp\left[ i \pi \oint_\Sigma 
(a^\ee \cup A^{\ee'} + a^\oo \cup A^{\oo' } ) \right] \\
 & = \sqrt{|H^1 (\Sigma , \bZ_2\times \bZ_2) | }~  \delta (A^{\ee' })\delta (A^{\oo'})\,,
\fe
where $\delta (A) =1$ if $A=0$, and $\delta(A)=0$ otherwise. 
The gauged theory is a 1+1d topological quantum field theory (TQFT) describing the spontaneous symmetry breaking (SSB) of $\bZ_2^{\ee'} \times \bZ_2^{\oo'}$ to nothing.  
Indeed, when $\Sigma = T^2$ and $A^{\ee'  } = A^{\oo' }=0$, $Z_\text{trivial}' =4$, which is the number of superselection sectors in a $\bZ_2^{\ee ' }\times \bZ_2^{\oo'}$ broken phase. 

On the other hand, the $\mathbf{S}$ gauging of  $Z_\text{cluster}$ gives back the original partition function:
\ie
~&Z' _\text{cluster} [A^{\ee' } ,A^{\oo' }]
= {1\over \sqrt{|H^1 (\Sigma , \bZ_2\times \bZ_2) | } }\sum_{(a^\ee, a^\oo) \in H^1 (\Sigma,\bZ_2\times\bZ_2)}  \\
&\exp\left[ i \pi \oint_\Sigma 
(  a^\ee \cup a^\oo +a^\ee \cup A^{\ee'} + a^\oo \cup A^{\oo' } ) \right] \\
 & =  \exp\left[ i \pi \oint_\Sigma 
A^{\ee'}\cup A^{\oo' }  \right] \,,
\fe
where the sum over $a^\ee$ sets $a^\oo = A^{\ee'}$. We conclude that
\ie
\mathbf{S}:~~~ \begin{array}{l} \text{trivial }\rightarrow \text{SSB}\\
\text{cluster}\rightarrow \text{cluster} \end{array}
\fe
This reproduces the lattice gauging in \eqref{app.S}. 
Since $Z_\text{cluster}$ is invariant under the $\mathbf{S}$ gauging, the half-gauging argument in \cite{Choi:2021kmx,Choi:2022zal} leads to a non-invertible global symmetry in the cluster SPT phase. (See \cite{Shao:2023gho} for a review of half-gauging.) 
In \cite{Seifnashri:2025fgd}, we will use gauging to obtain the duality defect in the cluster model; see also \cite{Sinha:2023hum,Seiberg:2024gek} for derivations of lattice defects using half-gauging.

\subsection{$\mathbf{TST}$ gauging in the continuum}

Next, we perform the twisted gauging labeled by $\mathbf{TST}$. 
Here $\mathbf{T}$ corresponds to stacking the partition function by the invertible field theory $Z_\text{cluster}$ in \eqref{Zcluster}:
\ie
\mathbf{T}:~~ Z[A^\ee ,A^\oo ] \mapsto \exp\left[i \pi \oint_\Sigma A^\ee \cup A^\oo\right ] Z[A^\ee ,A^\oo]\,.
\fe 
The $\mathbf{TST}$ gauging maps the trivial and the cluster theories to
\ie
\mathbf{TST}:~~~\begin{array}{l} \text{trivial }\rightarrow \text{trivial}\\
\text{cluster}\rightarrow \text{SSB} \end{array}
\fe
matching the lattice results in \eqref{TST}.

\section{Interface between the cluster and product states}\label{app:clusterprod}

We review the edge modes for the cluster model. 
For convenience of generalization to the case of non-invertible symmetries, we view the boundary condition for the cluster model as an interface between the cluster model and a trivial paramagnet.
More specifically, instead of working with an open chain, our space is a \textit{closed} chain of $L$ sites. The interface Hamiltonian is
\ie\label{appinterfaceH}
H_\text{interface}  =& -Z_1 X_2Z_3 -Z_2X_3Z_4 -\cdots - Z_{\ell-2} X_{\ell-1}Z_\ell\\
& - X_{\ell+1} -\cdots -X_L\,.  
\fe
Here $1< \ell < L$ and we assume  $\ell$ and $L$ to be both even for simplicity.  
In other words, we have the cluster Hamiltonian in the region $j=1, \cdots, \ell$ and the trivial paramagnet Hamiltonian in the region $j=\ell+1,\cdots, L$. 
The interface Hamiltonian enjoys the $\mathbb{Z}_2^\ee \times \mathbb{Z}_2^\oo$ symmetry generated by $\eta^\ee = \prod_{j=1, j :\text{even}}^L X_j, \eta^\oo = \prod_{j=1, j :\text{odd}}^LX_j$ on a closed chain. 
The advantage of this approach is that we never need to cut open the symmetry operators. 

A state $\ket{\psi}$ in the ground space of $H_\text{interface}$  satisfies 
\ie\label{clusterprod}
&Z_1 X_2 Z_3  = Z_2 X_3Z_4  =... = Z_{\ell-2} X_{\ell-1}Z_\ell =1\\
& X_{\ell+1} = ... = X_L =1\,.
\fe
Since there are only $L-2$ commuting Pauli constraints  for $L$ qubits, the ground space is 4-dimensional, signaling the edge modes at the interfaces. 

For any state $\ket{\psi}$ in the ground space, we can simplify the action of the $\mathbb{Z}_2^\ee \times\mathbb{Z}_2^\oo$ symmetry using \eqref{clusterprod}. 
The global $\eta^\ee,\eta^\oo$ operators factorize into products of local operators on the two interfaces:
\ie
&\eta^\ee  \ket{\psi}  =\eta_\text{L} ^\ee \eta_\text{R}^\ee \ket{\psi}\,,~~~\eta_\text{L}^\ee = Z_1\,,~~~\eta_\text{R}^\ee = Z_{\ell-1}X_\ell\,,\\
&\eta^\oo  \ket{\psi}  =\eta_\text{L} ^\oo \eta_\text{R}^\oo \ket{\psi}\,,~~~\eta_\text{L}^\oo =X_1 Z_2\,,~~~\eta_\text{R}^\oo = Z_\ell\,.
\fe
Here L and R stand for the left and right interfaces around $j=1$ and $j=\ell$, respectively. 
While $\eta^\ee $ and $\eta^\oo$ act linearly on the ground space (i.e., $\eta^\ee \eta^\oo = \eta^\oo \eta^\ee$), their local factors obey a projective algebra:
\ie
& \eta^\oo_\text{L} \eta^\ee _\text{L} =- \eta^\ee _\text{L}  \eta^\oo_\text{L}\,,\\
& \eta^\oo_\text{R} \eta^\ee _\text{R} =- \eta^\ee _\text{R}  \eta^\oo_\text{R}\,.
\fe
Equivalently, the left and right local factors $\eta^\oo_\text{L,R}$ are charged under $\eta^\ee$, and vice versa. 
This local projective algebra signals the relative difference in their SPT phases between the cluster state and the product state.

\section{Action of $\D$ on the SPT states \label{app:D.action}}

In this appendix, we compute the action of the non-invertible symmetry $\D$ on various SPT and interface states. We begin with the three SPT states.

\subsection{$\D\ket{\text{cluster}} = 2\ket{\text{cluster}}$}

To find the action of $\D$ on the cluster state, we use the expression $\D  = T \D^\ee \D^\oo$, where
\ie
~&\D^\ee = e^{2\pi i L\over16}
{1+\eta^{\ee}\over\sqrt{2}} {1-i X_L \over \sqrt{2} } \cdots {1- i Z_4 Z_2\over \sqrt{2} }
{1-i X_{2}\over\sqrt{2}} \,,\\
&\D^\oo=  e^{2\pi i L\over16}
{1+\eta^{\oo}\over\sqrt{2}} {1- i Z_{1}Z_{L-1} \over \sqrt{2} } \cdots {1-i X_{3}\over\sqrt{2}} {1- i Z_3Z_1\over \sqrt{2} } \,.
\fe
Here we have used a different presentation of $\D^\oo$ than the one in \eqref{Do.De}.

We first find the action of $\D^\oo$ on the cluster states as
\ie
	{}& \D^\oo\ket{\text{cluster}} = \\
	&\sqrt{2} e^{2\pi i L\over16} {1- i X_2\over \sqrt{2} }  {1- i Z_2 Z_4\over \sqrt{2} }
 \cdots {1- i X_{L} \over \sqrt{2} } \ket{\text{cluster}}\,, \label{Do.cluster}
\fe
where we have  repeatedly used $ Z_{j-1}X_jZ_{j+1} \ket{\text{cluster}} = \ket{\text{cluster}}$ to move the local factors from the right to the left. 
We also used  $\eta^\oo \ket{\text{cluster}} = \ket{\text{cluster}}$, which follows   writing $\eta^\oo = \prod_{j:\text{odd}} Z_{j-1}X_jZ_{j+1}$. Using \eqref{Do.cluster}, we find
\ie
	\D^\ee \D^\oo \ket{\text{cluster}} &= \\
	& = 2 e^{2\pi i L\over8} (-i X_2)(-iX_4) \cdots (-iX_L) \ket{\text{cluster}} \\
	&= 2 \eta^\ee \ket{\text{cluster}} = 2 \ket{\text{cluster}} \,.
\fe
Putting everything together, we find
\ie
	\D \ket{\text{cluster}} = 2T \ket{\text{cluster}} = 2 \ket{\text{cluster}}\,,
\fe
where the eigenvalue 2 is the quantum dimension of $\D$. 
The equality $T \ket{\text{cluster}} = \ket{\text{cluster}}$ follows from \eqref{cluster.state} and the facts that $T\V = \V T$ and $T \ket{++\cdots +} =  \ket{++\cdots +}$.

\subsection{$\D\ket{\text{odd}} = 2(-1)^{L/4}\ket{\text{odd}}$ \label{app:D.odd}}

We begin with the relation \eqref{circuit.o.d}, which we repeat here for convenience. Namely, 
\ie
	\ket{\text{odd}} = U_\text{odd} \ket{\text{cluster}}\,, \label{app:odd.u.cluster}
\fe
where
\ie
	U_\text{odd} = \prod_{n=1}^{L/2} Z_{2n} \, e^{\frac{\pi i}{4} (Z_{2n-1}Z_{2n+1}-1) } \,. \label{app:circuit.o.d}
\fe
We assume $L$ to be a multiple of 4 when discussing $\ket{\text{odd}}$, while other values of $L$ correspond to inserting  defects.

To find the action of $\D$ on the SPT state, we use the following commutation relation between $\D$ and $U_\text{odd}$:
\ie
	&\D \, U_\text{odd} = \left(\prod_{n= 1}^{L/2} e^{\frac{\pi i}{4} (X_{2n}-1)} \prod_{k=1}^{{L}/{4}} X_{ 4k-1} \right) \D \,, \\
	&= U_\text{odd} \left( \prod_{n= 1}^{\frac L2} e^{-\frac{\pi i}{4} (Z_{2n-1}Z_{2n+1}+X_{2n})}  \prod_{k=1}^{\frac{L}{4}} h_{ 4k-1} \right) \D, \label{app:action.u.d}
\fe
where 
\ie
h_j = Z_{j-1}X_jZ_{j+1}.
\fe
 Using this relation and the facts that $\D \ket{\text{cluster}} = 2 \ket{\text{cluster}}$ and $h_j \ket{\text{cluster}} = \ket{\text{cluster}}$, we find
\ie
	\D \ket{\text{odd}} &= U_\text{odd} \left( \prod_{n= 1}^{\frac L2} e^{-\frac{\pi i}{2} Z_{2n-1}Z_{2n+1}} \right) 2 \ket{\text{cluster}} \,, \\
	&= 2 {(-i)}^{\frac{L}{2}}  U_\text{odd}  \ket{\text{cluster}}  = 2 {(-1)}^{\frac{L}{4}} \ket{\text{odd}} \,.
\fe
Even though the model is defined for all $L=0$ mod 4, here we see an interesting dependence on $L$ mod 8. This is reminiscent of the mod 4 dependence in the $SO(3)$ Heisenberg chain discussed in \cite{Cheng:2022sgb}. We leave the study of this interesting sign for future investigations.

Finally, we have $\D\ket{\text{even}} = 2(-1)^{L/4}\ket{\text{even}}$, which follows from $T\D =\D T$ and $T \ket{\text{odd}} = \ket{\text{even}} $. 

\subsection{Action on the interface states}\label{app:noninv.interface}

Here, we find the action of the non-invertible symmetry on the edge modes localized on the interfaces between different SPT phases. We focus on the interface Hamiltonian $H_\text{cluster$|$odd}$ \eqref{interface} for the cluster and `odd' SPT states.

Define 
\ie
	X^{(\mathrm{L})} &= Y_{L-2} X_{L-1} Z_L\,,  ~ &X^{(\mathrm{R})} &= Z_\ell X_{\ell+1} Y_{\ell +2} \,,\\
	Z^{(\mathrm{L})} &= Z_{L-1}\,, &Z^{(\mathrm{R})} &= Z_{\ell+1}\,, \label{xlxr}
\fe
which are localized on the left and right interfaces and form a complete basis of operators acting on the ground space. 
We define the 4 interface ground states by $\ket{\varepsilon_{\mathrm{L}}, \varepsilon_{\mathrm{R}}}$, with $\varepsilon_\mathrm{L},\varepsilon_\mathrm{R}=\pm$, where they satisfy
\ie
	X^{(\mathrm{L})} \ket{\varepsilon_{\mathrm{L}}, \varepsilon_{\mathrm{R}}} &= \varepsilon_{\mathrm{L}} \ket{\varepsilon_{\mathrm{L}}, \varepsilon_{\mathrm{R}}} , &X^{(\mathrm{R})} \ket{\varepsilon_{\mathrm{L}}, \varepsilon_{\mathrm{R}}} &= \varepsilon_{\mathrm{R}} \ket{\varepsilon_{\mathrm{L}}, \varepsilon_{\mathrm{R}}} , \\
	Z^{(\mathrm{L})} \ket{\varepsilon_{\mathrm{L}}, \varepsilon_{\mathrm{R}}} &= \ket{-\varepsilon_{\mathrm{L}}, \varepsilon_{\mathrm{R}}} , &Z^{(\mathrm{R})} \ket{\varepsilon_{\mathrm{L}}, \varepsilon_{\mathrm{R}}} &= \ket{\varepsilon_{\mathrm{L}}, -\varepsilon_{\mathrm{R}}} ,
\fe
and are stabilized by:
\ie
	h_j=1 \quad &\text{ for } j = 1, \cdots, \ell \,, \\
	-h_{2n} = 1 \quad &\text{ for } n = \ell/2 +1, \cdots ,L/2 \,, \\
	Y_{2n}X_{2n+1}Y_{2n+2} = 1 \quad &\text{ for } n = \ell/2 + 1, \cdots ,L/2 - 2 \,. \label{appstabilizers}
\fe
The goal of this appendix is to find the action of the non-invertible symmetry on the ground space. We start with the $\bZ_2^\ee \times \bZ_2^\oo$ symmetry.

We first rewrite the $\bZ_2^\ee \times \bZ_2^\oo$ symmetry operators as products of the generators of the stabilizer \eqref{appstabilizers}:
\ie
	\eta^\ee &= \prod_{n=1}^{\ell / 2} h_{2n} \prod_{n=\ell/2 + 1}^{L /2} (-h_{2n}) \,, \\
	\eta^\oo &=  X^{(\mathrm{L})}  X^{(\mathrm{R})} \prod_{n=1}^{\frac \ell 2} h_{2n-1} \prod_{n=\frac \ell 2 + 2}^{\frac L 2 - 1} \left( Y_{2n-2}X_{2n-1}Y_{2n} \right)\,.
\fe
(Recall $L-\ell=0$ mod $4$ and $L$ is even.)
We have
\ie
	\eta^\ee \ket{\varepsilon_{\mathrm{L}}, \varepsilon_{\mathrm{R}}} &= \ket{\varepsilon_{\mathrm{L}}, \varepsilon_{\mathrm{R}}} , \\
	\eta^\oo \ket{\varepsilon_{\mathrm{L}}, \varepsilon_{\mathrm{R}}} &= X^{(\mathrm{L})}  X^{(\mathrm{R})} \ket{\varepsilon_{\mathrm{L}}, \varepsilon_{\mathrm{R}}} .
\fe

Now we study the action of $\D$. 
Since $ X^{(\mathrm{L})} \D = - \D X^{(\mathrm{L})} (-h_{L-2}) $ and $ X^{(\mathrm{R})} \D = - \D X^{(\mathrm{R})} (-h_{\ell + 2}) $, we find
\ie
	X^{(\mathrm{L})} \D \ket{\varepsilon_{\mathrm{L}}, \varepsilon_{\mathrm{R}}} &= - \D X^{(\mathrm{L})}\ket{\varepsilon_{\mathrm{L}}, \varepsilon_{\mathrm{R}}} \,,\\
	X^{(\mathrm{R})} \D \ket{\varepsilon_{\mathrm{L}}, \varepsilon_{\mathrm{R}}} &= - \D X^{(\mathrm{R})}\ket{\varepsilon_{\mathrm{L}}, \varepsilon_{\mathrm{R}}} \,. \label{D.X}
\fe
Moreover, $Z^{(\mathrm{L})} Z^{(\mathrm{R})} \D = - \D Z^{(\mathrm{L})} Z^{(\mathrm{R})} \prod_{n=\ell/2 + 1}^{L/2 -1} (-h_{2n})$,
which leads to 
\ie
	Z^{(\mathrm{L})} Z^{(\mathrm{R})} \D \ket{\varepsilon_{\mathrm{L}}, \varepsilon_{\mathrm{R}}} &= - \D Z^{(\mathrm{L})} Z^{(\mathrm{R})}\ket{\varepsilon_{\mathrm{L}}, \varepsilon_{\mathrm{R}}} \,. \label{D.ZZ}
\fe
Finally, since $\D^2 = (1+\eta^\ee)(1+\eta^\oo)$ we find
\ie
	\D^2 \ket{\varepsilon_{\mathrm{L}}, \varepsilon_{\mathrm{R}}} = 2\left(1+X^{(\mathrm{L})}X^{(\mathrm{R})}\right) \ket{\varepsilon_{\mathrm{L}}, \varepsilon_{\mathrm{R}}} \,. \label{DD}
\fe
Putting equations \eqref{D.X}, \eqref{D.ZZ}, and \eqref{DD} together, we find the action of $\D$ on the ground space ${\cal H}_{\text{cluster}|\text{odd}}$, up to a sign $\epsilon=\pm$, to be:
\ie
	\D \ket{\varepsilon_{\mathrm{L}}, \varepsilon_{\mathrm{R}}} =   i {\epsilon}   Z^{(\mathrm{L})} Z^{(\mathrm{R})} \left(X^{(\mathrm{L})} + X^{(\mathrm{R})}\right) \ket{\varepsilon_{\mathrm{L}}, \varepsilon_{\mathrm{R}}} \,. \label{D.up.to.sign}
\fe
To see this, we first note that the most general operator constructed from \eqref{xlxr} that anti-commutes with $X^{(\mathrm{L})}$, $X^{(\mathrm{R})}$, and $Z^{(\mathrm{L})} Z^{(\mathrm{R})} $, have the form $Z^{(\mathrm{L})} Z^{(\mathrm{R})} ( \alpha X^{(\mathrm{L})} + \beta X^{(\mathrm{R})})$ for some $\alpha,\beta \in \mathbb{C}$. Further requiring that this operator squares to $2 (1+X^{(\mathrm{L})}X^{(\mathrm{R})})$, leads to $\alpha = \beta = \pm i$. Thus we have found the action of $\D$ up to a  sign $\epsilon$. To fix $\epsilon$, below we compute the action of $\D$ on the interface state $\ket{+,+}$. The method is similar to the one we used in Appendix \ref{app:D.odd}.

We note that the interface state $\ket{+,+}$ can be prepared by acting with a finite-depth circuit on  $\ket{\text{cluster}}$ on a closed chain. Namely, we have
\ie
	\ket{+,+} = U^{++} \ket{\text{cluster}} ,
\fe
where
\ie
	U^{++} = \prod_{n=\ell/2 + 1}^{L/2} Z_{2n} \prod_{n=\ell/2 + 1}^{L/2-1} e^{-\frac{\pi i}{4} Z_{2n-1}Z_{2n+1}} \,.
\fe
The circuit $U^{++}$ is a  truncation of the circuit in \eqref{app:circuit.o.d} to the region between sites $\ell$ and $L$. 
One can verify this by studying the action of the circuit above on the operators $h_j = Z_{j-1} X_j Z_{j+1}$, which stabilize the cluster state. Indeed, the circuit $U^{++}$ acts as
\ie
	h_{2n}  &\mapsto \begin{cases}
	h_{2n}  & \text{for } 1 \leq n \leq \frac \ell 2 \\
	- h_{2n}  & \text{for } \frac \ell2+1 \leq n \leq \frac L2 \\
	\end{cases} \\
	h_{2n-1}  &\mapsto \begin{cases}
	h_{2n-1}  &  1 \leq n \leq \frac{\ell}{2} \\
	( - h_{\ell+2} ) X^{(\mathrm{R})} & n = \frac{\ell}{2}+1 \\
	h_{2n-2}(Y_{2n-2}X_{2n-1}Y_{2n})h_{2n} & \frac{\ell+4}{2} \leq n \leq \frac{L-2}{2} \\
	(-h_{L-2}) X^{(\mathrm{L})} & n = \frac L2
	\end{cases}
\fe

To study the action of $\D$ on the interface, we first find the action of $\D$ on $U^{++}$ as:
\ie
	&\D \, U^{++} = \left(\prod_{n=\ell/2 + 1}^{L/2-1} e^{-\frac{\pi i}{4} X_{2n}} \prod_{k=1}^{\frac{L-\ell}{4}} X_{\ell + 4k-1} \right) \D \,, \\
	&= U^{++} \left( \prod_{n=\frac{\ell}{2} + 1}^{\frac L2-1} e^{\frac{\pi i}{4} (Z_{2n-1}Z_{2n+1}+X_{2n})}  \prod_{k=1}^{\frac{L-\ell}{4}} h_{\ell + 4k-1} \right) \D.
\fe
Using this relation and the facts that $\D \ket{\text{cluster}} = 2 \ket{\text{cluster}}$ and $h_j \ket{\text{cluster}} = \ket{\text{cluster}}$, we find
\ie
	\D \ket{+,+} &= U^{++} \left( \prod_{n=\frac{\ell}{2} + 1}^{\frac L2-1} e^{\frac{\pi i}{2} Z_{2n-1}Z_{2n+1}} \right) 2 \ket{\text{cluster}} \,, \\
	&= 2 \left( {i}^{\frac{L-\ell}{2}-1} \, Z_{\ell+1}Z_{L-1} \right) U^{++}  \ket{\text{cluster}}  \,. \\
	&= -2i {(-1)}^{\frac{L-\ell}{4}} Z_{\ell+1}Z_{L-1} \ket{+,+} \,. \label{D++}
\fe
Hence, $\epsilon = - (-1)^{ L-\ell\over4}$.

Similarly, for the $\ket{-,-}$ state we have $\ket{-,-} = U^{--} \ket{\text{cluster}}$, where
\ie
	U^{--} = \prod_{n=\ell/2 + 1}^{L/2} Z_{2n} \prod_{n=\ell/2 + 1}^{L/2-1} e^{+\frac{\pi i}{4} Z_{2n-1}Z_{2n+1}} \,.
\fe
This relation leads to
\ie
	\D \ket{-,-} = +2i {(-1)}^{\frac{L-\ell}{4}} Z_{\ell+1}Z_{L-1} \ket{-,-} \,. \label{D--}
\fe

Equations \eqref{D++} and \eqref{D--} are compatible with \eqref{D.up.to.sign} and fix the action of $\D$ on the whole ground space. 
In summary, we find
\ie
	\D \ket{\varepsilon_{\mathrm{L}}, \varepsilon_{\mathrm{R}}} &= (-1)^{\frac{L-\ell}{4}} \left( Y^{(\mathrm{L})} Z^{(\mathrm{R})}  + Z^{(\mathrm{L})} Y^{(\mathrm{R})}  \right) \ket{\varepsilon_{\mathrm{L}}, \varepsilon_{\mathrm{R}}}  , \\
		\eta^\oo \ket{\varepsilon_{\mathrm{L}}, \varepsilon_{\mathrm{R}}} &= X^{(\mathrm{L})} X^{(\mathrm{R})}  \ket{\varepsilon_{\mathrm{L}}, \varepsilon_{\mathrm{R}}} , \\
	\eta^\ee \ket{\varepsilon_{\mathrm{L}}, \varepsilon_{\mathrm{R}}} &= \ket{\varepsilon_{\mathrm{L}}, \varepsilon_{\mathrm{R}}} .
\fe

\section{More on the absence of a Rep(D$_8$)-symmetric entangler \label{app:no.entangler}}

Here, we fill in the steps in the argument of the  non-existence of a  Rep(D$_8$) symmetric entangler in Section \ref{sec:entangler}. There we claimed that the existence of an entangler $\mathsf{U}$ satisfying \eqref{before.gauging} 
implies the existence of the unitary operator $\hat{\mathsf{U}}$ satisfying \eqref{after.gauging}. 
Below we provide an explicit argument for it.

The idea is to show that \eqref{after.gauging} is the image of \eqref{before.gauging} under the KT transformation. To simplify the discussion in this appendix, we identify the hatted operators with the unhatted operators. 
That is,  $\hat{X}_j = X_j $ and $\hat{Z}_j = Z_j$, which implies $\hat{\V} = \V$, $\hat\eta^\ee = \eta^\ee$, and $\hat\eta^\oo = \eta^\oo$. 
Under this  identification, the KT transformation is implemented by the non-invertible operator $\V\D\V$ as in \eqref{VDV}.

In this simplified notation, we will show that
\ie
	\mathsf{U} \D &= \D \mathsf{U} \,, & ~~~ \mathsf{U} \eta^\ee &=  \eta^\ee \mathsf{U} \,, \\
	\mathsf{U} \ket{\text{cluster}} &= \ket{\text{odd}} \,, &\mathsf{U} \eta^\oo &=  \eta^\oo \mathsf{U} \,, \label{app:before.gauging} 
\fe
implies the existence of the unitary operator $\hat{\mathsf{U}}=\hat{ \mathsf{U}}(X_j,Z_j)$ satisfying
\ie
	\hat{\mathsf{U}} \V &= \V \hat{\mathsf{U}} \,, & ~~~ \hat{\mathsf{U}} {\eta}^\ee &=  \eta^\ee \hat{\mathsf{U}} \,, \\
	\hat{\mathsf{U}} \ket{\widehat{\text{cluster}}} &= \ket{\widehat{\text{odd}}} \,, &\hat{\mathsf{U}} \eta^\oo &= \eta^\oo \hat{\mathsf{U}} \,. \label{app:after.gauging} 
\fe
Here, $\ket{\widehat{\text{cluster}}}$ and $\ket{\widehat{\text{odd}}}$ are, respectively, the \emph{unique} ground states of $\hat H_\text{cluster}$ and $\hat H_\text{odd}$ with eigenvalue $+1$ under $\eta^\ee,\eta^\oo,\V$. For instance, using the convention used around \eqref{DV.cluster}, we have
\ie
	\ket{\widehat{\text{cluster}}} = \frac{1}{2} \big(\ket{00} + \ket{01} + \ket{10} + \ket{11} \big)\,.
\fe
In infinite volume, each of these four short-range entangled states  $\ket{z_\ee z_\oo}$ becomes a superselection sector that  preserves $\bZ_2^\V$ and spontaneously breaks $\bZ_2^\ee \times \bZ_2^\oo$. 
Similarly, $\ket{\widehat{\text{odd}}}$ is the symmetric superposition of the 4 short-range entangled ground states of $\hat H_\text{odd}$. 
In infinite volume, each of these short-range entangled states becomes a superselection sector that only preserves $\text{diag}(\bZ_2^\ee \times \bZ_2^ \V)$. These two spontaneous symmetry breaking patterns are distinguished by the following order parameters:
\ie
	\bra{\widehat{\text{cluster}}} Z_j Z_{j+2n}\ket{\widehat{\text{cluster}}} = 1 \,, \quad \text{for $j=0,1$}\,,
\fe
and
\ie
	&\bra{\widehat{\text{odd}}} \left( Y_{0} \frac{1-Z_{-1}Z_{1}}{2} \right) \left( Y_{4k} \frac{1-Z_{4k-1}Z_{4k+1}}{2} \right) \ket{\widehat{\text{odd}}} = 1 \,, \\
	&\bra{\widehat{\text{odd}}} Z_1 Z_{4k+1} \ket{\widehat{\text{odd}}} = 1 \,.
\fe

Moreover, we will show that the action of KT transformation on $\mathsf{U}, \D, \ket{\text{cluster}}, \ket{\text{odd}}$ is given by $\hat{\mathsf{U}}, \V, \ket{\widehat{\text{cluster}}}, \ket{\widehat{\text{odd}}}$, in the sense that
\ie
	\left( \V\D \V \right) \mathsf{U} &= \hat{\mathsf{U}} \left( \V\D \V \right) , & \V\D \V \ket{\text{cluster}} &= 2 \ket{\widehat{\text{cluster}}} , \\
	 \left( \V\D \V \right) \D &= 2\V \left( \V\D \V \right) , & \V\D \V \ket{\text{odd}} &= 2 \ket{\widehat{\text{odd}}} . \label{app:KT}
\fe
The relation $( \V\D \V ) \D = 2\V ( \V\D \V )$ is derived in Appendix \ref{D.after.KT}, and $ \V\D \V \ket{\text{cluster}} = 2\ket{\widehat{\text{cluster}}}$ follows from equation \eqref{VDVD.cluster} of Appendix \ref{D.after.KT}.

\textbf{Existence of $\bZ_2^\ee \times \bZ_2^\oo$ symmetric $\hat{\mathsf{U}}$:} To argue for the existence of a $\bZ_2^\ee \times \bZ_2^\oo$ symmetric $\hat{\mathsf{U}}$, we note that the invariance of $\mathsf{U}$ under $\D$ means that the operator $\mathsf{U}$ remains invariant under the untwisted gauging $\mathbf{S}$. This implies that there is no mixed anomaly involving $\mathsf{U}$ and ${\bZ}_2^\ee \times {\bZ}_2^\oo$. Therefore, the image of  $\mathsf{U}$ under $\mathbf{TST}$ / KT transformation is an order 2, invertible, $\bZ_2^\ee \times \bZ_2^\oo$ symmetric operator $\hat{\mathsf{U}}$, which satisfies
\ie
	\left( \V\D \V \right) \mathsf{U} = \hat{\mathsf{U}} \left( \V\D \V \right). \label{image.of.u}
\fe
We note that \eqref{image.of.u} fixes the operator $\hat{\mathsf{U}}$ up to a factor of $\eta^\ee$ and $\eta^\oo$. This is because multiplying \eqref{image.of.u} from the right by $\V\D\V$ implies $(\V\D\V) \mathsf{U} (\V\D\V) = \hat{\mathsf{U}} (1+\eta^\ee)(1+\eta^\oo)$. In other words, if $\hat{\mathsf{U}}$ satisfies \eqref{image.of.u} (and \eqref{app:after.gauging}) then $\hat{\mathsf{U}}\eta^\ee$, $\hat{\mathsf{U}}\eta^\oo$, and $\hat{\mathsf{U}}\eta^\ee \eta^\oo$ also satisfy those relations.

\textbf{Action of KT on $\ket{{\text{odd}}}$:} The relation
\ie
	(\V\D \V) \, H_\text{odd} &= \hat H_\text{odd} \, (\V\D \V)
\fe
implies that the $\eta^\ee=\eta^\oo = 1$ spectrum of $H_\text{odd}$ and $\hat H_\text{odd}$ are related by the operator $\V\D \V$. 
In particular, $\ket{{\text{odd}}}$ and $\ket{\widehat{\text{odd}}}$ are the unique ground states of these Hamiltonians with eigenvalues $\eta^\ee=\eta^\oo = 1$. Hence, they are related by the operator $\V\D \V$:
\ie
	\V\D \V \ket{\text{odd}} = 2\ket{\widehat{\text{odd}}}.
\fe
The overall normalization is fixed by requiring the states to be normalized.

\textbf{Deriving \eqref{app:after.gauging}:} Now that we have verified \eqref{app:KT}, we are ready to derive \eqref{app:after.gauging}. Applying $\V \D \V$ to $\mathsf{U} \ket{\text{cluster}} = \ket{\text{odd}}$, and using \eqref{app:KT} we find
\ie
	\hat{\mathsf{U}} \left( \V\D \V \right) \ket{\text{cluster}} &= \left( \V\D \V \right) \ket{\text{odd}}\,,  \\
	\Rightarrow 2 \, \hat{\mathsf{U}} \ket{\widehat{\text{cluster}}} &= 2\ket{\widehat{\text{odd}}} \,. \label{app:u.hat.map}
\fe
Thus, we have shown that $\hat{\mathsf{U}}$ maps the ground states of $\hat{H}_\text{cluster}$ to those of $\hat H_\text{odd}$.

It remains to show that $\hat{\mathsf{U}}$ commutes with $\V$. 
Intuitively, this is follows from the fact that $\mathsf{U}$ and $\D$ commute, and their images under the KT transformation are $\hat{\mathsf{U}}$ and $\V$, respectively. In the following, we show this algebraically.
 
From $\mathsf{U} \D = \D\mathsf{U}$ and \eqref{app:KT} we find
\ie
	\hat{\mathsf{U}} \V  \left( \V\D \V \right) = \V\hat{\mathsf{U}}  \left( \V\D \V \right) \,.
\fe
By multiplying this equation on the right-hand-side by $\V\D \V$ we find
\ie
	\hat{\mathsf{U}} \V (1+\eta^\ee)(1+\eta^\oo) = \V\hat{\mathsf{U}}  (1+\eta^\ee)(1+\eta^\oo) \,. \label{uv.vu}
\fe
A priori, as we show below, this is only possible if
\ie
	\hat{\mathsf{U}} \V \hat{\mathsf{U}}^{-1} = \V, \, \V \eta^\ee, \, \V \eta^\oo, \, \V \eta^\ee \eta^\oo  \,. \label{uvuv}
\fe
\textbf{Proof of \eqref{uvuv}:} 
Define  
\ie
\mathsf{W} = \hat{\mathsf{U}} \V \hat{\mathsf{U}}^{-1} \V^{-1}\,.
\fe
Since we assume the entangler $\mathsf{U}$ to be locality-preserving, so is its image $\hat{\mathsf{U}}$, and therefore also $\mathsf{W}$. 
Next, since $\mathsf{W}$ commutes with $\eta^\ee,\eta^\oo$, it can be diagonalized to  the block-diagonal form $\text{diag}(\mathsf{W}_{++},\mathsf{W}_{+-},\mathsf{W}_{-+},\mathsf{W}_{--})$, where $\mathsf{W}_{\epsilon^\ee \epsilon^\oo}$ is the restriction of $\mathsf{W}$ to the eigenspace $\mathcal{H}_{\epsilon^\ee\epsilon^\oo}$ with $\eta^\ee = \epsilon^\ee$ and $\eta^\oo = \epsilon^\oo$. 
Then \eqref{uv.vu} implies $\mathsf{W}_{++} = 1$. 
Below we show that this implies that $\mathsf W$ acts trivially on all $\bZ_2^\ee \times\bZ_2^\oo$ symmetric local \textit{operators}.

Since $\mathsf{W}$ acts trivially on the states in $\mathcal{H}_{++}$,  given any  local operator $O_j$ that commutes with $\eta^\ee,\eta^\oo$, we have
\ie
	O_j \ket{\psi} = \mathsf{W} O_j \mathsf{W}^{-1} \ket{\psi} \,, \quad \text{for all } \ket{\psi} \in \mathcal{H}_{++}\,.
\fe
Given that $O_j$ and $\mathsf{W} O_j \mathsf{W}^{-1}$ are localized around site $j$, we can insert $Z_{j'}$'s far away from site $j$ and conclude that the relation above holds not only in $\mathcal{H}_{++}$, but on the entire Hilbert space. 
Hence, $O_j = \mathsf{W} O_j \mathsf{W}^{-1}$ for any $\bZ_2^\ee \times \bZ_2^\oo$ symmetric local operator $O_j$.

Next,  consider the local operators $O_j$ that satisfy $\eta^\ee O_j \eta^\ee = \epsilon^\ee O_j$ and $\eta^\oo O_j \eta^\oo = \epsilon^\oo O_j$ for some fixed $\bZ_2^\ee\times\bZ_2^\oo$ charges $\epsilon^\ee,\epsilon^\oo = \pm$. Multiplying two such operators, $O_j$ and $O'_{j'}$ that are well-separated, we find $O_j O'_{j'}$ that is $\bZ_2^\ee \times \bZ_2^\oo$ symmetric. Hence, it must commute with $\mathsf{W}$:
\ie
	(\mathsf{W} O_j \mathsf{W}^{-1})(\mathsf{W} O'_{j'} \mathsf{W}^{-1}) = O_j O'_{j'} \,.
\fe
Since $\mathsf{W} O_j \mathsf{W}^{-1}$ and $\mathsf{W} O'_{j'} \mathsf{W}^{-1}$ are well-separated, the equality above is only possible if $\mathsf{W} O_j \mathsf{W}^{-1} = \alpha(\epsilon^\ee,\epsilon^\oo) O_j$ and $\mathsf{W} O'_{j'} \mathsf{W}^{-1} = {\alpha({\epsilon^\ee,\epsilon^\oo})} O'_{j'}$ for $\alpha(\epsilon^\ee,\epsilon^\oo) = \pm$ that only depends on $\epsilon^\ee$ and $\epsilon^\oo$. Furthermore, $\alpha(\epsilon^\ee,\epsilon^\oo)$ is a bicharacter, i.e., $\alpha(+,+)=+$ and $\alpha(-,-)=\alpha(+,-)\alpha(-,+)$. This implies that the action of
\ie
	\mathsf{W} \qquad \text{and} \qquad \left(\eta^\ee\right)^{\frac{1-\alpha(-,+)}{2}} \left(\eta^\oo\right)^{\frac{1-\alpha(+,-)}{2}}
\fe
on local operators is the same. We have thus proved \eqref{uvuv}.

Finally, we show that only the first option in \eqref{uvuv} is compatible with anomalies. The last option is not possible since $\V$ is anomaly free but $\V \eta^\ee \eta^\oo $ is anomalous. To rule out the second possibility, we note that $\hat{\mathsf{U}} \V \hat{\mathsf{U}}^{-1} = \V\eta^\ee$ implies $\hat{\mathsf{U}} \V \eta^\oo \hat{\mathsf{U}}^{-1} = \V\eta^\ee \eta^\oo$, which is not possible since $\V \eta^\oo$ is anomaly free but $\V\eta^\ee \eta^\oo$ is anomalous. Similarly, we can also rule out the third possibility.  We conclude that $\hat{\mathsf{U}}$ must commute with $\V$. 
This completes the argument for \eqref{app:after.gauging} if there were a symmetric entangler, which subsequently led to the contradiction in Section \ref{sec:entangler}.

\section{Boundary conditions for the cluster model}

In this appendix we discuss symmetric boundary conditions for the cluster model.

The standard boundary condition for the cluster Hamiltonian on an open chain with $\ell$ sites (with even $\ell$) is:
\ie
H_\text{open}  = - \sum_{j=1}^{\ell-2} Z_j X_{j+1}  Z_{j+2} \,,
\fe
(This can be obtained from \eqref{appinterfaceH}  by fixing  the qubits on $j=\ell+1 ,\cdots , L$ to be all $\ket{+}$.) 
This Hamiltonian is  invariant under the symmetry operators
\ie\label{etaopen}
\eta^{\text{e}}_\text{open} = \prod_{j=2} ^\ell X_j\,,~~~
\eta^{\text{o}}_\text{open} = \prod_{j=1} ^{\ell-1} X_j\,.
\fe
The ground space is 4-dimensional, with each boundary being  2-fold degenerate. The Hamiltonian $H_\text{open}$ represents a $\bZ_2^\ee \times \bZ_2^\oo$ symmetric boundary condition with edge modes.

However, in  a general quantum system (such as  a continuum field theory or a lattice model without a tensor product Hilbert space), there may not be a canonical way to cut open the symmetry operator $U(g)$  of a global symmetry $G$ at the boundary.  
The notion of a ``symmetric boundary" therefore requires clarifications. 
One definition is the following\footnote{The different notions of symmetric boundary conditions for non-invertible symmetries are discussed in \cite{Choi:2023xjw}, where such a  boundary condition is referred to as a weakly symmetric boundary condition.} 
\begin{itemize}
\item A pair of boundary conditions are characterized by a Hamiltonian $H_\text{open}$ on an interval. The local terms of this Hamiltonian away from the boundaries are identical to the local terms in the Hamiltonian on a closed chain. 
\item There exist conserved operators $U_\text{open}(g)$ with $g\in G$ on an interval that commute with $H_\text{open}$ and realize the symmetry $G$ linearly, i.e., $U_\text{open}(g_1) U_\text{open}(g_2) =U_\text{open}(g_1g_2)$. The local factors of $ U_\text{open}(g)$ away from the boundaries are identical to the local factors of the conserved operator $U(g)$ on a closed chain.
\end{itemize}
Note that in this definition, the local factors of $ U_\text{open}(g)$ near the boundaries might be different from the local factors in the interior.

 It is generally expected \cite{Jensen:2017eof,Numasawa:2017crf,Thorngren:2020yht,Choi:2023xjw}   that given a quantum system with an ordinary global symmetry $G$ free of any 't Hooft anomaly (i.e., $G$ can be gauged and is compatible with a trivially gapped phase), there should exist a such symmetric boundary condition with no edge mode.

Since the $\bZ_2^\ee \times \bZ_2^\oo$ symmetry of the cluster model is free of 't Hooft anomalies, we expect there to be such a $\bZ_2^\ee \times \bZ_2^\oo$-symmetric boundary condition in the above sense. 
This is achieved by  considering a different Hamiltonian on the open chain:
\ie
\bar H_\text{open}  =  -  X_1 Z_2 - \left(\sum_{j=1}^{\ell-2} Z_j X_{j+1}  Z_{j+2}  \right) - Z_{\ell-1}X_\ell\,. 
\fe
Since the Hamiltonian consists of $L$ commuting Pauli operators, the ground space is non-degenerate. 
In particular, there is no local degeneracy at the boundary. 
This Hamiltonian on an open chain is also invariant under a $\mathbb{Z}_2^\ee \times \mathbb{Z}_2^\oo$ symmetry, generated by
\ie
\bar\eta^{\text{e}}_\text{open} =Z_1  \prod_{j=2} ^\ell X_j\,,~~~
\bar\eta^{\text{o}}_\text{open} =\left( \prod_{j=1} ^{\ell-1} X_j \right)Z_\ell\,,
\fe
which obey
 $[  \bar\eta^{\text{e}}_\text{open}  , \bar H_\text{open} ] = 0\,,~~~
[  \bar\eta^{\text{o}}_\text{open}  , \bar H_\text{open} ] = 0$. 

Let us summarize some features of this Hamiltonian $\bar H_\text{open}$:
\begin{itemize}
\item The local terms (i.e., $Z_j X_{j+1}Z_{j+2}$) in $\bar H_\text{open}$ in the interior of the chain are identical to the cluster Hamiltonian on a closed chain.
\item The local terms (i.e., $X_j$) of the $\mathbb{Z}_2^\ee \times \mathbb{Z}_2^\oo$ symmetry operators  $\bar \eta^\ee_\text{open},\bar\eta^\oo_\text{open}$ in the interior of the chain are identical to $\eta^\ee,\eta^\oo$ on a closed chain. 
\end{itemize} 
In this sense, $\bar H_\text{open}$ represents a $\mathbb{Z}_2^\ee \times \mathbb{Z}_2^\oo$ symmetric boundary condition for the cluster model with no edge mode.

This is not in contradiction with the usual statement in the literature that any symmetric boundary condition of the cluster model must have edge modes. This statement  assumes   the symmetry to take a strict on-site form in \eqref{etaopen}.  
However, this requirement does not have a generalization to non-onsite symmetries (such as our non-invertible symmetry $\D$), or to continuum field theories. 
Here we relax this requirement by allowing a local modification of the symmetry operators near the boundaries.

\bibliography{ref}

\end{document}